\documentclass[12pt]{article}
\usepackage{epsfig}
\def\beq{\begin{equation}}
\def\eeq{\end{equation}}
\def\beqa{\begin{eqnarray}}
\def\eeqa{\end{eqnarray}}
\newlength{\dinwidth}
\newlength{\dinmargin}
\newlength{\extraspace}
\newlength{\extraspaces}
\newcommand{\be}{\begin{equation}
\addtolength{\abovedisplayskip}{\extraspaces}
\addtolength{\belowdisplayskip}{\extraspaces}
\addtolength{\abovedisplayshortskip}{\extraspace}
\addtolength{\belowdisplayshortskip}{\extraspace}}
\newcommand{\ee}{\end{equation}}
\newcommand{\bdm}{\begin{displaymath}
\addtolength{\abovedisplayskip}{\extraspaces}
\addtolength{\belowdisplayskip}{\extraspaces}
\addtolength{\abovedisplayshortskip}{\extraspace}
\addtolength{\belowdisplayshortskip}{\extraspace}}
\newcommand{\edm}{\end{displaymath}}
\renewcommand{\thefootnote}{\fnsymbol{footnote}}
\def\simlt{\mathrel{\lower2.5pt\vbox{\lineskip=0pt\baselineskip=0pt
           \hbox{$<$}\hbox{$\sim$}}}}
\def\simgt{\mathrel{\lower2.5pt\vbox{\lineskip=0pt\baselineskip=0pt
           \hbox{$>$}\hbox{$\sim$}}}}
\newcommand{\ls}[1]
   {\dimen0=\fontdimen6\the\font
    \lineskip=#1\dimen0
    \advance\lineskip.5\fontdimen5\the\font
    \advance\lineskip-\dimen0
    \lineskiplimit=.9\lineskip
    \baselineskip=\lineskip
    \advance\baselineskip\dimen0
    \normallineskip\lineskip
    \normallineskiplimit\lineskiplimit
    \normalbaselineskip\baselineskip
    \ignorespaces}

\catcode`@=11
\newcount\@tempcntc
\def\@citex[#1]#2{\if@filesw\immediate\write\@auxout{\string\citation{#2}}\fi
  \@tempcnta\z@\@tempcntb\m@ne\def\@citea{}\@cite{\@for\@citeb:=#2\do
    {\@ifundefined
       {b@\@citeb}{\@citeo\@tempcntb\m@ne\@citea\def\@citea{,}{\bf ?}\@warning
       {Citation `\@citeb' on page \thepage \space undefined}}%
    {\setbox\z@\hbox{\global\@tempcntc0\csname b@\@citeb\endcsname\relax}%
     \ifnum\@tempcntc=\z@ \@citeo\@tempcntb\m@ne
       \@citea\def\@citea{,}\hbox{\csname b@\@citeb\endcsname}%
     \else
      \advance\@tempcntb\@ne
      \ifnum\@tempcntb=\@tempcntc
      \else\advance\@tempcntb\m@ne\@citeo
      \@tempcnta\@tempcntc\@tempcntb\@tempcntc\fi\fi}}\@citeo}{#1}}
\def\@citeo{\ifnum\@tempcnta>\@tempcntb\else\@citea\def\@citea{,}%
  \ifnum\@tempcnta=\@tempcntb\the\@tempcnta\else
   {\advance\@tempcnta\@ne\ifnum\@tempcnta=\@tempcntb \else \def\@citea{--}\fi
    \advance\@tempcnta\m@ne\the\@tempcnta\@citea\the\@tempcntb}\fi\fi}
\catcode`@=12

\begin{document}
\setcounter{footnote}{1}
\begin{flushright}
\today \\
LBNL-49074\\
SLAC-PUB-9057\\
UH-511-957-01
\end{flushright}
\begin{center}
\Large{{\bf Rare Charm Decays in the Standard Model and Beyond}}
\end{center}
\vspace{5mm}
\begin{center}
{
Gustavo Burdman$^a$, Eugene Golowich$^b$, JoAnne Hewett$^c$ 
and Sandip Pakvasa$^d$}\\
\vskip0.5cm
{\normalsize\it $^a$Lawrence Berkeley National Laboratory}\\
{\normalsize\it Berkeley, CA 94720}\\
\vskip0.2cm
{\normalsize\it $^b$Physics Department, University of
Massachusetts, }\\
{\normalsize\it Amherst, MA 01003}\\
\vskip0.2cm
{\normalsize\it $^c$Stanford Linear Accelerator Center, Stanford, CA
94309}\\
\vskip0.2cm
{\normalsize\it $^d$Department of Physics and Astronomy, University of 
Hawaii, }\\
{\normalsize\it Honolulu, HI 96822}
\end{center}
\vspace{0.30cm}
\thispagestyle{empty}
\begin{abstract}
We perform a comprehensive study of a number of rare charm decays,
incorporating the first evaluation of the QCD corrections to the short
distance contributions, as well as examining the long range effects.
For processes mediated by the $c\to u\ell^+\ell^-$ transitions,
we show that sensitivity to short distance physics exists in kinematic regions
away from the vector meson resonances that dominate the total rate.
In particular, we find that $D\to\pi\ell^+\ell^-$ and  
$D\to\rho\ell^+\ell^-$ are sensitive to non-universal soft-breaking
effects in the Minimal Supersymmetric Standard Model with R-parity 
conservation.  We separately study the sensitivity of 
these modes to R-parity violating effects and derive new bounds on 
R-parity violating couplings. 
We also obtain predictions for these decays within 
extensions of the Standard Model, including
extensions of the Higgs, gauge and fermion sectors, as well as models
of dynamical electroweak symmetry breaking.  
\end{abstract}

\newpage

\renewcommand{\thefootnote}{\arabic{footnote}}
\setcounter{footnote}{0}
\setcounter{page}{1}
\section{Introduction} 
\vspace{-0.2cm}
The remarkable success of the Standard Model (SM) in describing all 
experimental information currently available suggests that the quest
for deviations from it should be directed either at higher energy scales
or at small effects in low energy observables. To the latter group belong
the sub-percent level precision measurements of electroweak observables
at LEP and SLD as well as the Tevatron experiments~\cite{lep1}. 
Tests of the SM through quantum corrections have proved to be a powerful
tool for probing the high energy scales possibly related to electroweak
symmetry breaking and the flavor problem. The absence of flavor changing
neutral currents (FCNC) at tree level in the SM implies that processes
involving these currents are a primary test of the quantum structure of 
the theory. 
Most of the attention on FCNC has been focused on processes involving
$K$ and $B$ mesons, such as $K^0-\bar{K}^0$ and $B^0_{d(s)}-\bar{B}^0_{d(s)}$
mixing and also on rare decays involving transitions such as 
$s\to d\ell^+\ell-$, $s\to d\nu\bar\nu$, $b\to s\gamma$, 
$b\to s\ell^+\ell^-$, {\it etc}.  

The analogous FCNC processes in the charm sector have received 
considerably less scrutiny. This is perhaps due to the fact that, 
on general grounds, the SM expectations are very small both for 
$D^0-\bar{D}^0$ mixing~\cite{ddbar,GP98,Falk:2001}
as well as for FCNC decays~\cite{bghp,greub,other_rad}.     
For instance, there are no large non-decoupling effects arising from 
a heavy fermion in the leading one-loop contributions.  This is 
in sharp contrast with $K$ and $B$ FCNC processes, which are 
affected by the presence of the top quark in loops.
In the SM, $D$ meson FCNC transitions involve the rather light down-quark
sector which translates into an efficient GIM cancellation.
In many cases, extensions of the SM may upset this suppression and give 
contributions sometimes orders of magnitude larger than the SM.
In this paper we wish to investigate this possibility. 
As a first step, 
and in order to establish the existence of a clean window for the observation
of new physics in a given observable in rare charm processes, we must 
compute the SM contribution to such quantities. 
This is of particular importance
in this case due to the presence of potentially large 
long-distance contributions
which are non-perturbative in essence and therefore non-calculable by 
analytical methods. In general the flavor structure of charm FCNC favors
the propagation of light-quark-states as intermediate states which, if
dominant, obscure the more interesting short distance contributions that
are the true test of the SM. This is the situation in  $D^0-\bar{D}^0$ 
mixing~\cite{ddbar,GP98,Falk:2001} and in the $c\to u\gamma$ 
transition~\cite{bghp}. In the case of mixing, although the long distance 
effects seem to dominate over the SM short distance contributions, it is 
still possible that there is a window of one or two 
orders of magnitude between
these and the current experimental limit~\cite{cleodd}; 
the predictions of numerous extensions of the SM lie in this 
window~\cite{hnelson}.
On the other hand, charm radiative decays are completely dominated
by non-perturbative physics and do not constitute a suitable test of the
short distance structure of the SM or its extensions.

In what follows we investigate the potential of rare charm decays 
to constrain extensions of the SM. 
With the exception of $D^0\to\gamma\gamma$, we 
shall concentrate on the non-radiative FCNC transitions such as
$c\to u\ell^+\ell^-$, 
$c\to u\nu\bar\nu$ which enter in decays like $D^0\to\mu^+\mu^-$, 
$D\to X_u\ell^+\ell^-$, $D\to X_u\nu\bar\nu$, {\it etc}.
We extensively consider supersymmetry by studying 
the Minimal Supersymmetric SM (MSSM) as well as supersymmetric
scenarios allowing R-parity violation. We find that rare charm 
decays are potentially good tests of the MSSM and also serve to 
constrain R-parity violating couplings in kinematic regions away from 
resonances.  In charged dilepton modes, this mostly means at 
{\em low} dilepton mass.
In general, we find that this kinematic region, corresponding to large
hadronic recoil, is the most sensitive for new physics searches.  

The $D\to V\ell^+\ell^-$ decays were studied  in 
Ref.~\cite{fajfer1} in the SM without QCD corrections.  More recently the 
$D\to\pi\ell^+\ell^-$ 
decays were examined in Ref.~\cite{fajfer2}
in the SM and some of its extensions, including the 
MSSM. We compare these predictions with ours, and find some discrepancies
in the SM calculation of the long distance contributions. 
We also emphasize the importance of $D\to V\ell^+\ell^-$ in the 
MSSM due to its enhanced sensitivity to the electromagnetic dipole
moment operator entering in $c\to u\gamma$.

In the next section we calculate the SM short distance contributions
including QCD corrections 
and estimate long distance effects for various decay modes. 
In Section~3 we study possible extensions of the SM that might produce  
signals which fall below current experimental limits but above the SM results
of Section~2. We summarize and conclude in Section~4.

As a final comment, we note the following convention and 
notation used throughout the paper. Many quantities relating 
to both SM and also new physics are chiral, involving 
projection operators for left-handed (LH) 
and right-handed (RH) massless fermions.  We shall employ the notation 
\begin{equation}
\Gamma_{{\rm L}, {\rm R}} \equiv {1 \pm \gamma_5 \over 2} \ , 
\qquad 
\Gamma_{{\rm L}, {\rm R}}^\mu 
\equiv {\gamma^\mu (1 \pm \gamma_5) \over 2} \\ 
\label{proj}
\end{equation}
for scalar projection operators $\Gamma_{\rm L,R}$ and 
vector projection operators $\Gamma_{\rm L,R}^\mu$.  
The chiral projections of fermion field $q$ are thus expressed as 
\begin{equation}
q_{{\rm L},{\rm R}} \ \equiv \ \Gamma_{{\rm L}, {\rm R}} ~q \ \ .
\label{chiral}
\end{equation}

\section{The Standard Model Contributions}
\label{sec:sm}
In this section we study the Standard Model contributions to various 
charm meson rare decays. At the time of this writing, there are 
no reported events of the type we are 
considering.  We group the decay modes by their common short distance
structure.  In each case we address both the 
perturbative short distance amplitude and the
effects of the non-perturbative long-range propagation of intermediate
hadronic states.  Due to the non-perturbative nature of the 
underlying physics, the long distance effects cannot be 
calculated with controlled uncertainties.  Therefore we find it 
prudent to generate estimates by using several distinct
approaches, such as vector meson dominance (VMD) for 
processes with photon emission and/or calculable 
unitarity contributions. In this way, we hope to obtain a 
reasonable measure of the uncertainty involved in the 
calculation, and at the same time, obtain bounds 
on the magnitude of long-distance contributions which are not 
overly model dependent.

\subsection{Meson Lepton-antilepton Transitions $D \to X \ell^+\ell^-$}

As we shall discuss, this mode is likely to be observed 
at forthcoming 
B and Charm factory/accelerator experiments.   
We start with the calculation of both short and long distance 
contributions to the inclusive rate. We then compute the rates for 
various exclusive modes.

\subsubsection{The Short Distance Contribution to $D \to X_u \ell^+ \ell^-$}  
The short distance contribution is induced at one loop in the SM.  
It is convenient to use an effective description with the 
$W$ boson and the b-quark
being integrated out as their thresholds are reached, 
respectively, in the renormalization group evolution~\cite{gw},
\begin{eqnarray}
{\cal H}_{\rm eff} & = & -{4G_F\over\sqrt 2} \left[ \sum_{q=d,s,b} 
C_1^{(q)}(\mu)O_1^{(q)}(\mu) + C_2^{(q)}(\mu)O_2^{(q)}(\mu) + 
\sum_{i=3}^{10} C_i(\mu)O_i(\mu)\right]  \,,\,\, m_b<\mu<M_W \nonumber\\
{\cal H}_{\rm eff} & = & -{4G_F\over\sqrt 2}\left[  \sum_{q=d,s} 
C_1^{(q)}(\mu)O_1^{(q)}(\mu) + C_2^{(q)}(\mu)O_2^{(q)}(\mu) + 
\sum_{i=3}^{10} C'_i(\mu)O'_i(\mu)\right]  \,,\,\, \mu<m_b\,, 
\label{heff}
\end{eqnarray}
with $\{O_i\}$ being the complete operator basis, 
$\{C_i\}$ the corresponding Wilson coefficients and $\mu$ the renormalization
scale; the primed quantities
indicate those where the b-quark has been eliminated.  Note that we must
keep all terms of order $1/M_W^2$ above the scale $\mu=m_b$ in this decay
as opposed to radiative decays.
In Eq.~(\ref{heff}), the Wilson coefficients contain the dependence
on the Cabibbo-Kobayashi-Maskawa (CKM) matrix elements $V_{qq'}$. 
As was pointed
out in Ref.~\cite{bghp}, the CKM structure of these transitions is drastically
different from that of the analogous $B$ meson processes. 
The operators $O_1$ and $O_2$ are explicitly split into their 
CKM components
\begin{equation}
O_1^{(q)}=(\bar{u}_L^{\alpha}\gamma_\mu q_L^{\beta})
(\bar{q}_L^{\beta}\gamma^\mu c_L^\alpha)\ , \qquad  
O_2^{(q)}=(\bar{u}_L^{\alpha}\gamma_\mu q_L^{\alpha})
(\bar{q}_L^{\beta}\gamma^\mu c_L^\beta)\ \ ,
\label{o1q}
\end{equation}
where $q=d,s,b$, and $\alpha$, $\beta$ are contracted color indices. 
The rest of the operator basis is defined in the standard way.
The QCD penguin operators are given by
\begin{eqnarray}
O_3&=&(\bar{u}_L^{\alpha}\gamma_\mu c_L^{\alpha})\sum_{q}(\bar{q}_L^\beta
\gamma^\mu q_L^\beta)\ , \qquad 
O_4 =(\bar{u}_L^{\alpha}\gamma_\mu c_L^{\beta})\sum_{q}(\bar{q}_L^\beta
\gamma^\mu q_L^\alpha)\ \ , \nonumber\\
O_5&=&(\bar{u}_L^{\alpha}\gamma_\mu c_L^{\alpha})\sum_{q}(\bar{q}_R^\beta
\gamma^\mu q_R^\beta)\ , \qquad 
O_6=(\bar{u}_L^{\alpha}\gamma_\mu c_L^{\beta})\sum_{q}(\bar{q}_R^\beta
\gamma^\mu q_R^\alpha)~~, \label{qcdpen}
\end{eqnarray}
the electromagnetic and chromomagnetic dipole operators are
\beq
O_7 = \frac{e}{16\pi^2}m_c(\bar{u}_L\sigma_{\mu\nu}c_R)F^{\mu\nu}\ ,
\qquad 
O_8 = \frac{g_s}{16\pi^2}m_c(\bar{u}_L\sigma_{\mu\nu}
T^a c_R)G^{\mu\nu}_a\ \ ,
\label{gdipole}
\eeq
and finally the four-fermion operators coupling directly to the charged leptons
are
\beq
O_9=\frac{e^2}{16\pi^2} (\bar{u}_L\gamma_\mu c_L)( \bar{\ell}\gamma^\mu
\ell) \ , \qquad 
O_{10}=\frac{e^2}{16\pi^2} (\bar{u}_L\gamma_\mu c_L)( \bar{\ell}\gamma^\mu
\gamma_5\ell) \ \ .
\label{laxial}
\eeq
The matching conditions at $\mu=M_W$ for the Wilson coefficients
of the operators $O_{1-6}$ are 
\beq
C_1^{q}(M_W) = 0\ , \qquad C_{3-6}(M_W)=0 \ , \qquad 
C_2^{q}(M_W)=-\lambda_q \ \ ,
\label{mco16}
\eeq
with $\lambda_q=V^*_{cq}V_{uq}$. The corresponding conditions for the 
coefficients of the operators $O_{7-10}$ are 
\begin{eqnarray}
C_7(M_W)&=&-\frac{1}{2}\left\{ \lambda_s F_2(x_s) +\lambda_b F_2(x_b)
\right\}~~,
\nonumber\\
C_8(M_W)&=&-\frac{1}{2}\left\{ \lambda_s D(x_s) +\lambda_b D(x_b)
\right\}~~, 
\nonumber\\
C^{(')}_9(M_W)&=&\sum_{i=s,(b)}\lambda_i 
\left[-\left(F_1(x_i)+2\bar{C}(x_i)\right)
+\frac{\bar{C}(x_i)}{2s_w^2}\right]~~,
\nonumber\\
C^{(')}_{10}(M_W)&=&-\sum_{i=s,(b)} \lambda_i\frac{\bar{C}(x_i)}{2s_w^2}~~.
\label{c10mw}
\end{eqnarray}
In Eqs.~(\ref{c10mw}) we define $x_i=m_i^2/M_W^2$, 
the functions $F_1(x)$, $F_2(x)$ and $\bar{C}(x)$ are 
those derived in Ref.~\cite{IL81} and 
the function $D(x)$ was defined in Ref.~\cite{bghp}. 

To compute the $c\to u\ell^+\ell^-$ rate at leading order, 
operators in addition to $O_7$, $O_9$ and $O_{10}$ must contribute. 
Even in the absence of the strong interactions, the insertion of the 
operators $O_2^{(q)}$ in a loop would give a contribution 
sometimes referred to as leading order mixing of $C_2$ with 
$C_9$. 
When the strong interactions are included, further mixing
of the four-quark operators with $O_{7-10}$ occurs. 
The effect of these QCD corrections in the renormalization group (RG)
running from $M_W$ down to $\mu=m_c$ is of 
particular importance in $C_7^{\rm eff}(m_c)$, 
the coefficient determining the $c\to u\gamma$ amplitude. As was shown in
Ref.~\cite{bghp}, the QCD-induced mixing with $O_2^{(q)}$
dominates $C_7^{\rm eff}(m_c)$. The fact that the main contribution
to the $c\to u\gamma$ amplitude comes from the insertion of four-quark 
operators inducing light-quark loops signals the presence of large long
distance effects. This was confirmed in Ref.~\cite{bghp} where these 
non-perturbative contributions were estimated and found to dominate the 
rate. Therefore, in the present calculation we will take into account 
effects of the strong interactions in $C_7^{\rm eff}(m_c)$. 
On the other hand, as mentioned above, the operator $O_9$ mixes
with four-quark operators even in the absence of QCD 
corrections~\cite{buras}. 
Finally, the RG running does not affect $O_{10}$, {\it i.e.} $C_{10}(m_c)=
C_{10}(M_W)$.  
Thus, in order to estimate the $c\to u\ell^+\ell^-$ amplitude it 
is a good approximation to consider the QCD effects only 
where they are dominant, {\it i.e.} in $C_7^{\rm eff}(m_c)$, whereas 
we expect these to be less dramatic in $C_9^{\rm eff}(m_c)$.

The leading order mixing of $O_2^{(q)}$ with $O_9$ results in 
\begin{equation}
C^{('){\rm~eff}}_9=C^{(')}_9(M_W)+\sum_{i=d,s,(b)}\lambda_i\left[
-\frac{2}{9}{\rm ~ln}\frac{m_i^2}{M_W^2} +\frac{8}{9}\frac{z_i^2}{\hat s}
-\frac{1}{9}\left(2+\frac{4z_i^2}{\hat s}\right)
\sqrt{\left| 1-\frac{4z_i^2}{\hat s}\right|}~{\cal T}(z_i) \right] \ , 
\label{c9eff}
\end{equation}
where we have defined
\begin{equation}
{\cal T}(z)=\left\{
\begin{array}{cc}
2{\rm ~arctan}\left[\frac{1}{\sqrt{\frac{4z^2}{\hat s}-1}}\right] 
& ({\rm for~}\hat s < 4 z^2) \\ 
\\
{\rm ~ln}\left|\frac{1+\sqrt{1-\frac{4z^2}{\hat s}}}
{1-\sqrt{1-\frac{4z^2}{\hat s}}}\right| -i\pi~ & ({\rm for~}
\hat s > 4 z^2) \ \ ,
\end{array}
\right.
\label{deft}
\end{equation}
and $\hat s\equiv s/m_c^2$, $z_i\equiv m_i/m_c$.
The logarithmic dependence on the internal quark mass $m_i$ in the 
second term of Eq.~(\ref{c9eff}) cancels against a similar term in 
the Inami-Lim function $F_1(x_i)$ entering in $C_9(M_W)$, 
leaving no spurious divergences in the $m_i\to 0$ limit. 

To compute the differential decay rate in terms of the Wilson coefficients, 
we use the two-loop QCD corrected value of $C_7^{\rm eff}(m_c)$ as obtained 
in Ref.~\cite{greub}, compute $C_9^{\rm eff}(m_c)$ from Eq.~(\ref{c9eff}), and 
$C_{10}(m_c)=C_{10}(M_W)$ from Eq.~(\ref{c10mw}). The 
differential decay rate in the approximation of 
massless leptons is given by
\begin{eqnarray}
\frac{d\Gamma_{c\to u\ell^+\ell^-}}{d\hat s}&=& 
\tau_D~\frac{G_F^2\alpha^2m_c^6}{768\pi^5} ~(1-\hat s)^2 
\left[ 
\left(\left|C_9^{('){\rm~eff}}(m_c)\right|^2+\left|C_{10}\right|^2\right)
\left(1+2\hat s\right)\right.\nonumber\\
& &\left.+ 12 ~C_7^{\rm eff}(m_c){\rm ~Re}\left[C_9^{('){\rm~eff}}(m_c)\right]
+ 4 \left(1+\frac{2}{\hat s}\right)\left|C_7^{\rm eff}(m_c)\right|^2
\right]~~,
\label{dbs}
\end{eqnarray}
where $\tau_D$ refers to the lifetime of either $D^{\pm}$ or $D^0$. 
We estimate the inclusive branching ratios for $m_c=1.5$~GeV, $m_s=0.15$~GeV, 
$m_b=4.8$~GeV and  $m_d=0$,
\beq
{\cal B}r_{D^+\to X_u^+ e^+e^-}^{\rm (sd)} \simeq 2\times10^{-8} \ ,
\qquad 
{\cal B}r_{D^0\to X_u^0 e^+e^-}^{\rm (sd)} \simeq 8\times10^{-9} \ \ .
\label{d0br}
\eeq
It is useful to observe that the dominant contributions to the 
rates in Eq.~(\ref{d0br}) come from the leading order mixing
of $O_9$ with the four-quark operators $O_2^{(q)}$, the second term in 
Eq.~(\ref{c9eff}).
As noted above, the dominance of light-quark
intermediate states in the short distance contributions is a signal of the 
presence of large long distance effects. However, when considering 
the contributions of various new physics scenarios, it should be kept in 
mind that their magnitudes must be compared to the mixing of these operators. 
Shifts in the matching conditions for the 
Wilson coefficients $C_7$, $C_9$ and $C_{10}$, even when large, are 
not enough to overwhelm the long distance effects
in most extensions of the SM. These considerations will be helpful
when we evaluate what type of new physics scenarios might be relevant in these 
decay modes. 

\subsubsection{The Long Distance Contributions to $D \to X_u \ell^+ \ell^-$}  
\label{longd_sec}
As a first estimate of the contributions of long distance physics we 
will consider the resonance process $D\to X V\to X\ell^+\ell^-$, where 
$V=\phi,\rho,\omega$. 
We isolate contributions from this particular 
mechanism by integrating $d\Gamma/dq^2$ over each resonance peak 
associated with an exchanged vector or pseudoscalar meson.
The branching ratios thus 
obtained (we refer to each such branching ratio 
as ${\cal B}r^{\rm (pole)}$)  are in the 
${\cal O}(10^{-6})$ range.  Modes experiencing the largest effects 
are displayed in Table~\ref{tab:res} (see also Ref.~\cite{sizh}), 
where we compare our theoretically derived 
branching ratios with existing experimental bounds~\cite{PDG00}. 
Due to the small $\eta \to \ell^+\ell^-$ and 
$\eta' \to \ell^+\ell^-$ branching ratios, the dominant 
contributions arise from $V^0$ exchange.

\begin{table*}[t]
\caption{Examples of $D \to P V^0 \to P \ell^+\ell^-$ Mechanism.\label{tab:res}}
\vspace{0.4cm}
\begin{center}
\begin{tabular}{c|c|c}
\hline \hline 
Mode & ${\cal B}r^{\rm (pole)}$ & ${\cal B}r^{\rm (expt)}$ 
\\ \hline 
 $D^+ \to \pi^+ \phi \to \pi^+ e^+ e^-$ & $1.8 \cdot 10^{-6}$  & 
 $< 5.2 \cdot 10^{-5}$   \\ 
 $D^+ \to \pi^+ \phi \to \pi^+ \mu^+ \mu^-$ & $1.5 \cdot 10^{-6}$  & 
 $< 1.5 \cdot 10^{-5}$   \\ 
 $D_s^+ \to \pi^+ \phi \to \pi^+ e^+ e^-$ & $1.1 \cdot 10^{-5}$  & 
 $< 2.7 \cdot 10^{-4}$   \\ 
 $D_s^+ \to \pi^+ \phi \to \pi^+ \mu^+ \mu^-$ & $0.9 \cdot 10^{-5}$  & 
 $< 1.4 \cdot 10^{-4}$   \\ 
\hline\hline
\end{tabular}
\end{center}
\end{table*}  

This result suggests that the long distance contributions
overwhelm the short distance physics and possibly any new physics
that might be present. 
However, as we will see below this is not always the case. 
A more thorough treatment requires looking at all the kinematically
available regions in $D\to X_u\ell^+\ell^-$, not just the resonance region.
In order to do this, 
the effect of these states can be 
thought of as a shift
in the short distance coefficient $C_9^{\rm eff}$ in Eq.~(\ref{c9eff}), 
since $V\to\ell^+\ell^-$ selects a vector coupling for the leptons. 
This follows from Ref.~\cite{lms}, which incorporates 
in a similar manner
the resonant contributions to $b\to q\ell^+\ell^-$ decays via a 
dispersion relation for $\ell^+\ell^-\to $~hadrons. This procedure
is manifestly gauge invariant. 
The new contribution can be written via the replacement~\cite{lms} 
\begin{equation}
C_9^{\rm eff}\to C_9^{\rm eff} + \frac{3\pi}{\alpha^2}
\sum_{i} 
\kappa_i\frac{m_{V_i}\Gamma_{V_i\to\ell^+\ell^-}}{m_{V_i}^2-s-im_{V_i}
\Gamma_{V_i}}~,
\label{c9res}
\end{equation}
where the sum is over the various relevant resonances, $m_{V_i}$ and 
$\Gamma_{V_i}$ are the resonance mass and width, and the factor
$\kappa_i \sim {\cal O}(1)$ is a free parameter adjusted to fit 
the non-leptonic decays
$D\to X V_i$ when the $V_i$ are on shell. We obtain $\kappa_\phi\simeq 3.6$,
$\kappa_\rho\simeq 0.7$  and $\kappa_\omega\simeq 3.1$. The last value 
comes from assuming ${\cal B}r_{D^+\to\pi^+\omega}=10^{-3}$, since a direct 
measurement is not available yet.

\begin{figure}
\hskip 3.0cm
\epsfig{figure=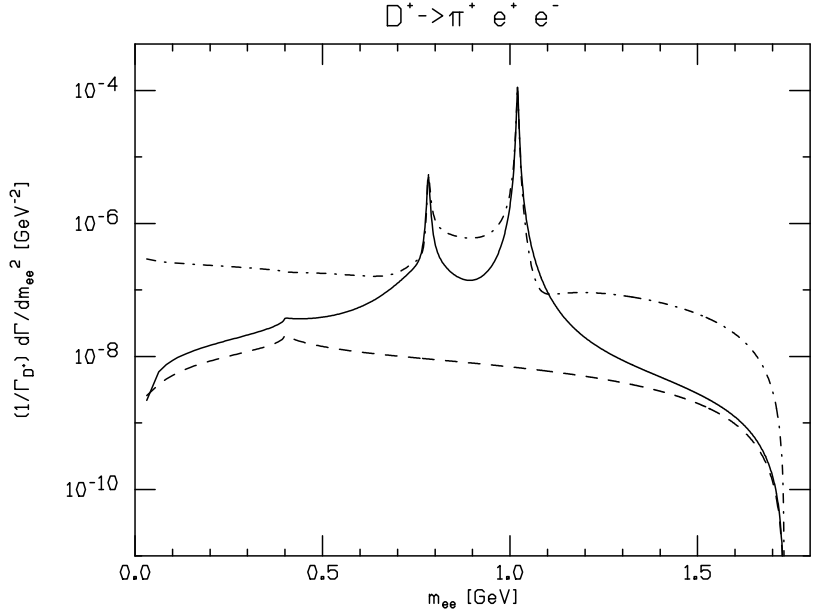,height=4.2in,angle=90}
\caption{The dilepton mass distribution for 
$D^+\to\pi^+ e^+e^-$, normalized to $\Gamma_{D^+}$. 
The solid line shows the sum of the short and the long distance 
SM contributions.
The dashed line corresponds to the short distance contribution only.
The dot-dash 
line includes the allowed R-parity violating contribution from Supersymmetry 
(see Section~\ref{rpvsection})}
\label{pill}
\end{figure}

As a first example we study the $D^+\to\pi^+e^+e^-$ decay. 
The main long-distance contributions come from 
the $\phi$, $\rho$ and $\omega$ resonances. The $\eta$ and $\eta'$
effects are negligibly small.
The dilepton mass distribution for this decay
takes the form
\begin{eqnarray}
\frac{d\Gamma}{ds} = \frac{G_F^2\alpha^2}{192\pi^5}
|{\bf p}_\pi|^3\,|f_+(s)|^2\,
\left( \left|\frac{2m_c}{m_D} C_7^{\rm eff} + C_9^{\rm eff}\right|^2
+|C_{10}|^2\right)~,
\label{meedist}
\end{eqnarray} 
where $s=m_{ee}^2$ is the squared of the dilepton mass. 
Here we have make use of the heavy quark spin symmetry relations
that relate the matrix elements of $O_7$ to the ``semileptonic''
matrix elements of $O_9$ and $O_{10}$~\cite{iw90}. 
An additional form-factor 
is formally still present, but its contribution to the 
decay rate is suppressed by $(m_\ell/m_D)^2$ and is neglected here.
For the form-factor $f_+(s)$ we make use of the prediction
of Chiral Perturbation Theory for Heavy Hadrons~\cite{chpthh}, which 
at low recoil gives
\begin{equation}
f_+(s) = \frac{f_D}{f_\pi}\frac{g_{D^*D\pi}}{(1-s/M_{D^{*}}^2)}
\label{fp}~,
\end{equation}
where we use the recent CLEO measurement~\cite{cleo_g} 
$g_{D^*D\pi}=0.59\pm0.1\pm0.07$, and we take $f_D=200$~MeV. 
In Fig.~\ref{pill} we present this distribution as a function of the 
dilepton mass. 
The two narrow peaks are the $\phi$ and the $\omega$, which sit on top
of the broader $\rho$.
The total rate results in ${\cal B}r_{D^+\to\pi^+e^+e^-}
\simeq 2\times10^{-6}$. Although  most of this branching ratio arises
from the intermediate $\pi^+\phi$ state, we can see from Figure~\ref{pill}
that new physics effects as low as $10^{-7}$ can be observed as long as 
such sensitivity is achieved in the regions away from the $\omega$ and $\phi $
resonances, both at low and high dilepton mass squared.

Similarly, we can consider the decay $D^+\to\rho^+e^+e^-$. 
Since there is less data available at the moment on the 
$D\to V V'$ modes, we will take the values of the $\kappa_i$ in 
Eq.~(\ref{c9res}) from the fits to the $D^+\to\pi^+ V$ case studied above.
For the semileptonic form-factors we use the extracted values from 
the $D\to K^*\ell\nu$ data~\cite{drhoff} and assuming $SU(3)$ 
symmetry\footnote{The $D\to\rho$ form-factors will be extracted with 
precision at Charm and B factories. In the meantime, we do not believe the 
assumption of $SU(3)$ symmetry will affect our main conclusions here.}. 
\begin{figure}
\hskip 3.0cm
\epsfig{figure=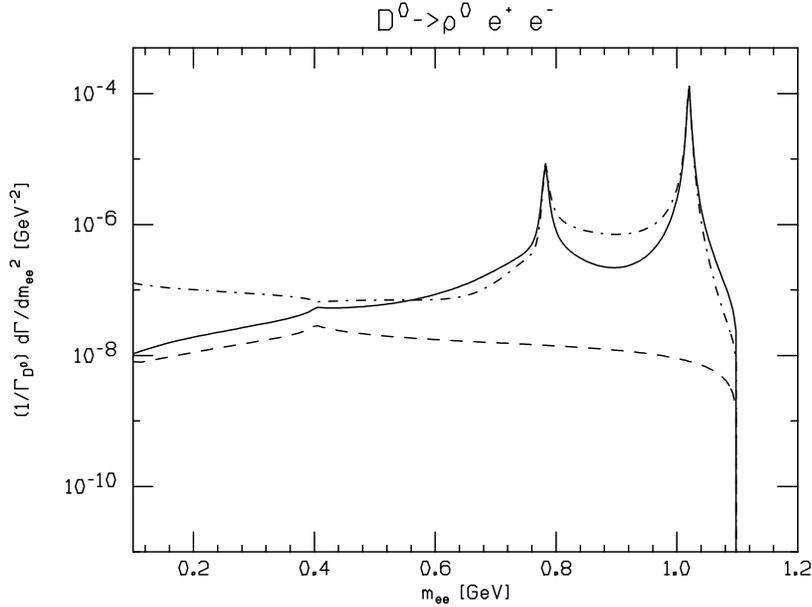,height=4.2in,angle=90}
\caption{The dilepton mass distribution for 
$D^0\to\rho^0 e^+e^-$, normalized to $\Gamma_{D^0}$. 
The solid line shows the sum of the short and the 
long distance SM contributions.
The dashed line corresponds to the short distance contribution only. The 
dot-dash 
line includes the allowed R-parity violating contribution from Supersymmetry 
(see Section~\ref{rpvsection})} 
\label{rholl}
\end{figure}
The total integrated branching ratio is 
${\cal B}r_{D^0\to\rho^0e^+e^-}=1.8\times 10^{-6}$ ({\it i.e.}
${\cal B}r_{D^+\to\rho^+e^+e^-}=4.5
\times 10^{-6}$). As can be seen in Fig.~\ref{rholl}, once 
again most of this rate 
comes from the resonance contributions.
However, there is also a region -in this 
case confined to low values of $m_{ee}$ due to the kinematics- 
where sensitive measurements 
could test the SM short distance structure of these transitions.
In addition, the $\rho$ modes contain angular information 
in the form of a forward-backward asymmetry for the lepton pair. 
Since this asymmetry arises as a consequence of the interference between 
the vector and the axial-vector couplings of the leptons, 
it is negligible in the SM since the vector couplings due to vector mesons
overwhelm the axial-vector couplings. This is true even away from the 
resonance region, partly because of the large width of the $\rho$ and
partly since the coefficient $C_9^{('){\rm eff}}$ and 
$C_7^{('){\rm eff}}$ get large enhancements due to mixing with 
$O_2$ and from the 
QCD corrections, whereas $C_{10}$ -the axial-vector coupling- 
is not affected by any of these.  This results in a very small interference.
We expand on this point and consider the possibility of large asymmetries
from physics beyond the SM in Section~\ref{rpvsection}. 
For both the $\pi$ and $\rho$ modes 
the sensitivity to new physics effects is reserved 
to large ${\cal O}(1)$ enhancements
since the long distance contributions are still important even when 
away from the resonances. 

We finally compare our results in Figs.~\ref{pill} and~\ref{rholl}
with those obtained in Refs.~\cite{fajfer1} and~\cite{fajfer2}.  
The short distance calculations in both these papers do not include 
the tree-level mixing of $O_9$ with $O_2$. This effect determines 
most of the short distance amplitude. Also, as mentioned above, 
this piece cancels the logarithm in Eq.~(\ref{c9eff}), a scheme
dependent term of no physical significance. If this cancellation did not
take place the logarithm would be the largest contribution to $C_9$.  
In addition, in Ref~\cite{fajfer1} the QCD corrections are not included. 
We also differ in the long distance results, which 
dominate these decays.
\noindent
For $D\to\pi\ell^+\ell^-$ the authors of Ref.~\cite{fajfer1} make use
of the factorization approximation, as well as heavy hadron chiral perturbation
theory for both pseudoscalars and vector mesons. It is far from 
clear that the use of both  approximations in $D$ decays is warranted.
\noindent
For the case of $D\to\rho\ell^+\ell^-$, the results of Ref.~\cite{fajfer2}
show a large enhancement at low $q^2$ when compared with Fig.~\ref{rholl}. 
However, a $1/q^2$ enhancement can only appear as a result of 
non-factorizable contributions. 
This is clear from Ref.~\cite{gp95} and~\cite{soares}:  
the factorization amplitude for $D\to\rho V$, 
when combined with a gauge invariant $(\gamma-V)$ mixing, leads to a null
contribution to $D\to V\ell^+\ell^-$. This is due to the 
fact that the mixing of the operator $O_2$ with $O_7$ 
is non-factorizable~\cite{soares}. 
A resonant contribution to $O_7$, leading to a $1/q^2$ 
behavior, is then proportional to $C_7^{\rm eff}$, which is 
mostly given by the 
$O_2$ mixing. In addition, when compared with the 
usual short distance matrix element of $O_7$, this resonant contribution
will be further suppressed by the factor 
$g_V(q^2) A^{\rm nf}(q^2)~$,
where $g_V(q^2)$ is the $(\gamma-V)$ mixing form-factor, and 
$A^{\rm nf}(q^2)$ parametrizes the non-factorizable 
amplitude $\langle \rho V|O_7|D\rangle$, which is 
of ${\cal O}(\Lambda_{\rm QCD}/m_c)$~\cite{qcdfac}. 
Thus, even if we take the 
on-shell values for these quantities, the resonant contribution
to $O_7$ is likely to be below $10\%$ of the SM short distance contribution.
The actual off-shell values at low $q^2$ far from the resonances are likely
to be even smaller. 
We then conclude that the $1/q^2$ enhancement is mostly 
given by the short distance contribution. This is only noticeable
at extremely small values of the dilepton mass, so that it is
likely to be beyond the experimental sensitivity in the electron 
modes (due to Dalitz conversion), whereas in the muon modes 
it lies beyond the physical region.
On the other hand, the factorizable pieces contribute to the 
matrix elements of $O_9$, just as in Eq.~(\ref{c9res}), 
and give no enhancement at low values of $q^2$. 

\subsection{Neutrino-antineutrino Emission $D \to P \nu_\ell {\bar\nu}_\ell$}

In the Standard Model, decays such as  
\beqa
D^+ (p) \to \pi^+ (p') ~ \nu_\ell (k) ~ {\bar \nu}_\ell ({\bar k}) 
\qquad {\rm and} \qquad 
D^0 (p) \to {\bar K}^0 (p') ~ \nu_\ell (k) ~ {\bar \nu}_\ell ({\bar k}) 
\label{2nu1}
\eeqa
will have branching ratios which are generally (but, as we shall 
show, not always) too small to measure.  Such decays thus represent 
attractive modes for new physics searches.  

\subsubsection{The Short Distance Contribution 
$c \to u \nu_\ell {\bar \nu}_\ell$}
These decay modes are induced by 
$Z$ penguin as well as box diagrams.
The corresponding effective hamiltonian takes the form
\begin{equation}
{\cal H}_{\rm eff}=\frac{G_F}{\sqrt{2}}\frac{\alpha}{2\pi s^2_W}
\sum_{\ell=e,\mu,\tau} \left\{ \lambda_s X^\ell(x_s) + \lambda_b X^\ell(x_b)
\right\} (\bar{u}_L\gamma_\mu c_L)(\bar{\nu}_L^\ell \gamma^\mu \nu_L^\ell)~.
\label{hnunu}
\end{equation}
The functions 
in Eq.~(\ref{hnunu}) are defined by 
$X^\ell(x_i)=\bar{D}(x_i,m_\ell)/2$, 
with the functions $\bar D$ given in Ref.~\cite{IL81}.
Although we  have explicitly kept the dependence on the charged lepton masses 
arising from 
the box diagrams, this is of numerical significance only when considering
the strange quark contributions 
with an internal tau lepton. In any case, the branching ratios in the 
SM are unobservably small. For instance, one has
\beq 
{\cal B}r_{D^+\to X_u \nu\bar\nu}^{\rm (s.d.)} \simeq
1.2\times10^{-15} \ , \qquad 
{\cal B}r_{D^0\to X_u \nu\bar\nu}^{\rm (s.d.)} \simeq 
5.0\times10^{-16} \ \ ,
\label{dpnn}
\eeq
where the contributions of all neutrinos have been included. 

\begin{figure}
\vskip .1cm
\hskip 1.9cm
\epsfig{figure=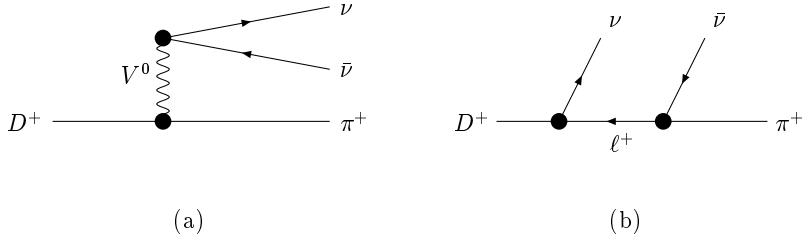,height=1.2in}
\caption{Some long distance contributions. \hfill 
\label{fig:dggfig6}}
\end{figure}

\subsubsection{Long Distance Contributions to $D \to P \nu_\ell 
{\bar \nu}_\ell$}

Long-distance contributions to the exclusive transition 
$D \to P \nu_\ell {\bar \nu}_\ell$ ($P$ is a pseudoscalar meson) can have 
just hadrons, just leptons or both hadrons and leptons 
in the intermediate state.  Examples of the first two 
cases are depicted respectively in 
Fig.~\ref{fig:dggfig6}(a) and Fig.~\ref{fig:dggfig6}(b).  

As a simple model of the purely hadronic intermediate state, 
we consider in detail the 
non-leptonic weak process $D (p)\to \pi (p') V^0 (q)$  
followed by the conversion $V^0(q) \to \nu_\ell (k) 
{\bar \nu}_\ell ({\bar k})$, 
{\it cf} Fig.~\ref{fig:dggfig6}(a).  We determine first the 
$V^0 \to \nu_\ell {\bar \nu}_\ell$ ($V^0 = \phi, \rho^0, \omega$) 
vertex, which has the invariant amplitude
\begin{equation}
{\cal M}_{V^0  \to \nu_\ell  {\bar \nu}_\ell} \simeq  
\left({g_2 \over 2 \cos\theta_w} \right)^2 
{1 \over M_Z^2}~ {\bar u}(k) \Gamma_\mu^{\rm L} 
v ({\bar k}) \ \langle 0 | \ \sum_q J^\mu_q \ | V^0 \rangle \ \ ,
\label{2b2a}
\end{equation}
where $J^\mu_q$ is the current coupling quark $q$ to the $Z$ 
gauge boson.  Only the vector part of the current contributes 
and we find  
\begin{equation}
{\cal M}_{V^0 \to \nu_\ell {\bar \nu}_\ell} \simeq  {2 G_F \over \sqrt{2}} 
h_V {\bar u}(k) \epsilon_V^\mu \Gamma_\mu^{\rm L} 
v ({\bar k}) \ \ .
\label{2b2b}
\end{equation}
Using the measured electromagnetic transitions 
$V^0 \to \ell^+ \ell^-$ \ ($V^0 = \rho^0,\omega,\phi$) 
as input, we find for the coupling $h_V$ 
\beq
|h_V| = 
\left\{
\begin{array}{ll}
(3/2 - 2 s_w^2)M_\phi^2/f_\phi  \simeq 0.112~{\rm GeV}^2 & (V = \phi) \\
(9/8 - 2 s_w^2)M_\rho^2/f_\rho - 3 M_\omega^2/8f_\omega 
\simeq 0.107~{\rm GeV}^2  & (V = \rho) \\
-(9/8 - 2 s_w^2)M_\omega^2/f_\omega + 3 M_\rho^2/8f_\rho 
\simeq 0.008~{\rm GeV}^2 & (V = \omega) \ \ ,
\end{array} \right.
\label{2b2c}
\eeq
where we adopt the numerical values of $f_\phi, f_\rho, f_\omega$ 
listed in Ref.~\cite{gp95}.

The corresponding transition amplitude 
for the non-leptonic $D$ decay process is then 
\begin{equation}
{\cal M}_{D \to P \nu_\ell {\bar \nu}_\ell}^{({\rm V}^{\rm 0})} = 
G_F^2 M_D^2 {1 \over q^2 - (M_V - i \Gamma_V/2)^2} 
F(q^2) h_V(q^2) {\bar u}(k) p' \cdot\gamma \Gamma_{\rm L} 
v ({\bar k}) \ \ ,
\label{2b2d}
\end{equation}
where $q \equiv p - p' = k + {\bar k}$ is the four-momentum carried 
by the virtual vector meson and $F(q^2)$ appears in the 
$D\to V^0 P$ amplitude.  We find for the $q^2$-distribution
\begin{equation}
{d\Gamma_{D \to P \nu_\ell {\bar \nu}_\ell} 
\over d q^2} = {G_F^4 M_D^4 \over 192 \pi^3} 
{|{\bf p}'| \over M_D^2} {F^2 (q^2) h_V^2 (q^2) \over
(q^2 - M_V^2)^2 + \Gamma_V^2 M_V^2} \left( 
 (q \cdot p')^2 - {q^2 M_V^2 \over 4} \right) \ .
\label{2b2e}
\end{equation}
We have used data from non-leptonic decays 
into pseudoscalar-vector final states ($D \to P + V^0$) to serve as input for 
$D^+ \to \pi^+ \nu_\ell {\bar \nu}_\ell$ ($\rho^0$ pole), 
$D^0 \to {\bar K}^0 \nu_\ell {\bar \nu}_\ell$ ($\rho^0, \omega, \phi$ poles) 
and $D_s^+ \to \pi^+ \nu_\ell {\bar \nu}_\ell$ ($\omega,\phi$ poles). 
Taking the largest contributor in each category, we obtain 
\beqa
{\cal B}r_{D^+ \to \pi^+ \nu {\bar \nu}} &\simeq& 
5.1 \times 10^{-16} \qquad (V = \rho^0) 
\nonumber \\
{\cal B}r_{D^0 \to {\bar K}^0 \nu {\bar \nu}} &\simeq& 
2.4 \times 10^{-13} \qquad (V = \phi) 
\nonumber \\
{\cal B}r_{D_s^+ \to \pi^+ \nu {\bar \nu}} &\simeq& 
7.8 \times 10^{-15} \qquad (V = \phi) \ \ ,
\label{2b2f}
\eeqa
where we have summed over the three neutrino flavors.
Although this analysis pertains to just the amplitudes of 
Fig.~\ref{fig:dggfig6}(a), we believe our results reflect 
the order of magnitude to be expected for other hadronic 
intermediate states as well.  All such processes lead to 
unmeasurably small branching ratios.

There will also be amplitudes with single lepton intermediate 
states, as in Fig.~\ref{fig:dggfig6}(b).  For electron and 
muon intermediate states, the amplitude for 
$D(p) \to P(p') \nu_\ell (k) {\bar \nu}_\ell ({\bar k})$
is reducible to 
\beq
{\cal M}_{D \to P \nu_{(e,\mu)} 
{\bar \nu}_{(e,\mu)}}^{\rm (lept.)} = 
- 2 G_F^2 V_{ud} V_{cd}^* {\bar u}(k) p \cdot\gamma \Gamma_{\rm L}
v ({\bar k}) + {\cal O}(m_{(e,\mu)}^2) \ \ .
\label{2b2g}
\eeq
These lead to the branching ratios 
\beq
{\cal B}r_{D^+ \to \pi^+ \nu_{(e,\mu)} {\bar \nu}_{(e,\mu)}} 
\simeq 1.8 \times 10^{-16} \ , \qquad 
{\cal B}r_{D_s^+ \to \pi^+ \nu_{(e,\mu)} {\bar \nu}_{(e,\mu)}} 
\simeq 3.8 \times 10^{-15} \ \ ,
\label{2b2h}
\eeq
which are again too small for detection.  

There remains the case in which $\tau^+$ propagates as 
the intermediate state.  This differs from the above 
cases involving $e$ and $\mu$ propagation in that for 
part of the $\nu_\tau$-${\bar \nu}_\tau$ phase space, 
the intermediate $\tau^+$ is on the mass shell.  The mode
$D_s^+ \to \tau^+ + \nu_\tau$ has been 
observed\footnote{In this experiment, only the leptonic 
decay mode $\tau^+ \to \ell \nu_\ell {\bar \nu}_\tau$ 
was detected.~\cite{L3}} with 
${\cal B}r_{D_s^+ \to \tau^+ + \nu_\tau} = (7 \pm 4)$\% 
whereas $D^+ \to \tau^+ + \nu_\tau$ has not (the predicted 
branching ratio is ${\cal B}r_{D^+ \to \tau^+ + \nu_\tau} 
\simeq 9.2 ~10^{-4}$).  Once the on-shell $\tau^+$ has 
been produced, its branching ratio to decay into a 
given meson can be appreciable, {\it e.g.} 
${\cal B}r_{\tau \to \rho^+ {\bar \nu}_\tau} 
\simeq 0.25$, ${\cal B}r_{\tau \to \pi^+ {\bar \nu}_\tau} 
\simeq 0.11$, {\it etc}.  Such transitions, although 
involving production of a $\nu{\bar \nu}$ pair in the 
final state, should be measurable at a $B$ and/or Charm factory.   

\subsection{Two Photon Emission $D^0 \to \gamma \gamma$}
The amplitude for the transition $D^0 (p) \to \gamma (q_1, \lambda_1) 
\gamma (q_2, \lambda_2)$ can be expressed as 
\begin{equation}
{\cal M}_{D^0\gamma\gamma} = \epsilon_\mu^\dagger (1) 
\epsilon_\nu^\dagger (2) 
\left[ (q_1^\nu q_2^\mu - q_1 \cdot q_2 ~ g^{\mu\nu} ) ~
C_{D^0 \gamma \gamma}  
+ i \epsilon^{\mu\nu\alpha\beta} q_{1\alpha} q_{2\beta} ~
B_{D^0 \gamma \gamma} \right]  \ \ .
\label{dgg1}
\end{equation}
The invariant amplitudes $B_{D^0 \gamma \gamma}$ and  
$C_{D^0 \gamma \gamma}$ are
P-conserving and P-violating, respectively, and carry 
units of inverse energy.  They contribute 
to the $D^0 \to \gamma \gamma$ branching ratio as 
\beq
{\cal B}r_{D^0 \to \gamma\gamma} 
= {M_D^3 \tau_{D^0} \over 64 \pi } \left[ \ 
| B_{D^0 \gamma \gamma} |^2 + 
 |C_{D^0 \gamma \gamma} |^2 \ \right] \ \ .
\label{dgg2}
\eeq
The amplitude in Eq.~(\ref{dgg1}) is sometimes written in 
the equivalent form
\begin{equation}
{\cal M}_{D^0 \gamma\gamma} = {C_{D^0 \gamma \gamma} \over 2} 
F_1^{\mu\nu} F_{2\mu\nu} + i 
{B_{D^0 \gamma \gamma} \over 2} 
F_{1\mu\nu} {\tilde F}_{2\mu\nu} \ \ , 
\label{dgg3}
\end{equation}
where $F^{\mu\nu} \equiv i ( q^\mu \epsilon^\nu - 
q^\nu \epsilon^\mu )$ and 
${\tilde F}^{\mu\nu} \equiv \epsilon^{\mu\nu\alpha\beta} 
F_{\alpha \beta}/2$.

\subsubsection{The Short Distance Contribution $c {\bar u} \to \gamma\gamma$}
Consider the quark level transition $c\to u 
\gamma \gamma$.  This can arise via one-particle 
irreducible (1PI) processes in which both photons arise 
from the interaction vertex or one-particle reducible (1PR) 
processes in which at least one of the photons is 
radiated from the initial state $c$-quark or final state 
$u$-quark.  

\begin{figure}
\vskip .1cm
\hskip 1.9cm
\epsfig{figure=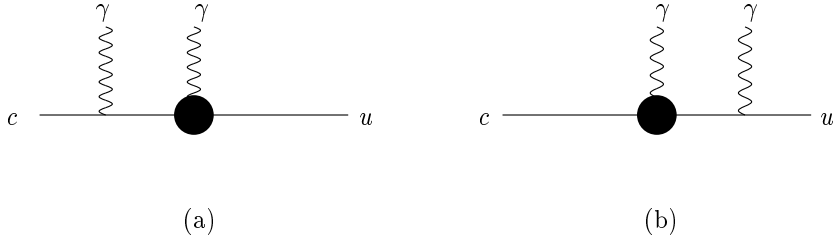,height=1.2in}
\caption{$1PR$ contributions to $c \to u \gamma\gamma$. \hfill 
\label{fig:dggfig1}}
\end{figure}

To estimate the $c\to u \gamma \gamma$ 
amplitude, we employ an approximation which 
makes use of known results on the related process 
$c \to u \gamma$. According to Ref.~\cite{greub}, 
the two-loop $c \to u \gamma$ vertex is
\begin{equation}
{\cal M}_{c u \gamma}^{\rm (s.d.)} = {4 G_F \over \sqrt{2}}
{e \over 16 \pi^2}
A m_c \sigma_{\mu\nu} \Gamma_{\rm R} F^{\mu\nu} \ \ ,
\end{equation}
where $|A| \simeq 0.0047$. Keeping in mind that there are 
additional diagrams which must be accounted for in a 
complete two-loop analysis, we shall use this as input to the 1PR graphs 
depicted in Fig.~\ref{fig:dggfig1}.  
The dominant contribution to the $c \to u \gamma \gamma$ 
amplitude involves photon emission 
from the $u$-quark.  
To ensure that the effect is indeed 'short-range', we 
follow the locality procedure employed in Ref.~\cite{yao}. 
This yields for $c {\bar u} \to \gamma\gamma$ the amplitude 
\begin{equation}
|B_{D^0\gamma\gamma}^{\rm (s.d.)}| = 
|C_{D^0\gamma\gamma}^{\rm (s.d.)}| = { G_F \alpha 
\over 3 \sqrt{2}\pi} 
{ m_c \over M_D - m_c} ~f_D ~|A| \ \ ,
\end{equation}
resulting in the branching ratio
\begin{equation}
{\cal B}r_{D^0\to\gamma\gamma}^{({\rm s.d.})} \ \simeq \ 3 \times 
10^{-11} \ \ ,
\label{brsd}
\end{equation}
for the choice $M_D - m_c \simeq 0.3$ GeV.  

\subsubsection{Long Distance Contributions to $D^0 \to \gamma\gamma$}

We shall model long-distance contributions to the $D^0 \to \gamma
\gamma$ amplitude using the vector meson dominance (VMD) 
mechanism and the unitarity constraint.  The latter can only 
be done in a limited context since there will be many unitarity 
contributions. We will consider several one-particle intermediate 
states (as used in $K \to \gamma\gamma$ decays) as well as 
the two-particle $K^+K^-$ intermediate state. 

\begin{figure}
\vskip .1cm
\hskip 4.6cm
\epsfig{figure=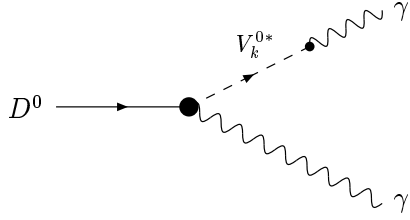,height=1.1in}
\caption{Vector dominance (VMD) contribution.}
\label{fig:dggfig2}
\end{figure}

\begin{center}
{\bf Vector Meson Dominance}
\end{center}


One can view ({\it c.f.} Fig.~\ref{fig:dggfig2})
the $D^0 \to \gamma \gamma$ amplitude as the 
single VMD process
\beq
D^0 \to \gamma \ + \ \sum_k \ ~V^{0*}_k 
\ \to \ \gamma + \gamma \ \ .
\label{vmd2}
\eeq
We have previously used the VMD mechanism to model 
the general single-photon emission $D \to M + \gamma$ ($M$ is some 
non-charm meson)~\cite{bghp}. It is straightforward to 
extend our analysis to the $D^0 \to \gamma \gamma$ mode, 
as long as care is taken in the $D^0 \to \gamma \gamma$ amplitude 
to ensure gauge invariance and Bose-Einstein statistics. 
The amplitudes used in the $D^0 (p) 
\to V^0 (k) + \gamma(q)$ transition are defined as
\begin{equation}
{\cal M}_{D V \gamma} = \epsilon_V^{\mu\dagger}(k,\lambda_V)
\epsilon_\gamma^{\nu\dagger}(q,\lambda_\gamma) \left[ 
C_V (k_\nu q_\mu - k \cdot q g_{\mu\nu}) + 
i B_V \epsilon_{\mu\nu\alpha\beta} k^\alpha q^\beta  \right] \ \ .
\label{vmd3}
\end{equation}
The VMD amplitude that we calculate is therefore of the form
\beqa
B_{D^0\gamma\gamma}^{\rm (vmd)} =  
\sum_i \ {2 e \over f_{V_i}}~ B_{V_i} ~\eta_i \ , \qquad 
C_{D^0\gamma\gamma}^{\rm (vmd)} 
= \sum_i \ {2 e \over f_{V_i}}~ C_{V_i}~ \eta_i \ \ ,
\label{vmd3a}
\eeqa
where $f_V$ is the coupling for the $V^0 - \gamma$ 
conversion amplitude, the index '$i$' 
refers to the specific vector meson 
($\rho^0, \omega^0, \phi^0$) 
and $\eta_i$ is a factor accounting for the VMD extrapolation 
made in $q^2$.  We take $\eta_i \simeq 1/2$ as a reasonable choice.

The values in Table~\ref{tab:vmd} are somewhat lower than 
those which would be obtained from the $V\gamma$ amplitudes in 
Ref.~\cite{bghp}.  The main reason for this is the central 
value for ${\cal B}r_{D^0 \to \phi \rho^0}$, which is a 
numerically significant input to the VMD calculation, 
cited in the Particle 
Data Group compilation has decreased by a factor of about three 
between 1994 and 2000.  Using the central values in Table~\ref{tab:vmd} 
and assuming positive interference between the various amplitudes 
to provide the maximal VMD signal gives the branching ratio 
\beq
{\cal B}r_{D^0 \to\gamma \gamma}^{\footnotesize{\rm (vmd)}} = 
\left( 3.5~^{+4.0}_{-2.6}  \right) \times 10^{-8} \ \ .
\label{vmd4}
\eeq

\begin{table*}[t]
\caption{VMD Amplitudes ($10^{-8}$ GeV$^{-1}$).\label{tab:vmd}}
\vspace{0.4cm}
\begin{center}
\begin{tabular}{c|c|c}
\hline \hline 
$D^0 \to V^0\gamma$ & $B_{D^0\gamma\gamma}^{\rm vmd}$ 
& $C_{D^0\gamma\gamma}^{\rm vmd}$ \\ \hline
$D^0 \to\rho^0 \gamma$ & $0.036~ (1 \pm 0.7)$ & $0.045 ~(1 \pm 0.3)$ \\
$D^0 \to \omega^0 \gamma$ & $0.011 ~(1 \pm 0.5)$ & $0.012 ~(1 \pm 0.5)$\\
$D^0 \to \phi^0 \gamma$ & $0.047 ~(1 \pm 0.7)$ &  $0.036 ~(1 \pm 0.4)$\\
\hline \hline
\end{tabular}
\end{center}
\end{table*}  

\pagebreak
\begin{center}
{\bf Single-particle Unitarity Contribution}  
\end{center}

\begin{figure}
\vskip .1cm
\hskip 4.6cm
\epsfig{figure=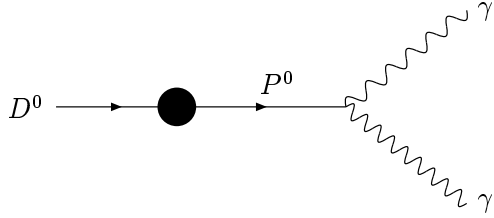,height=1.1in}
\caption{Weak mixing contribution.} \hfill 
\label{fig:dggfig4}
\end{figure}

In this category of amplitudes ({\it cf.} Fig.~\ref{fig:dggfig4}) 
the $D^0$ mixes with a spinless meson (either a pseudoscalar $P_n$
or a scalar $S_n$) and finally decays into a photon pair, 
\beqa
B_{D^0 \gamma\gamma}^{\rm (mix)}\  &=& \ \sum_{P_n} \ 
\langle P_n | {\cal H}_{wk}^{\rm (p.c.)} | D^0 \rangle 
~{1 \over M_D^2 - M_{P_n^2}} ~ B_{P_n \gamma\gamma}
\nonumber \\
C_{D^0 \gamma\gamma}^{\rm (mix)} \ &=&\  \sum_{S_n} \ 
\langle S_n | {\cal H}_{wk}^{\rm (p.v.)} | D^0 \rangle 
~{1 \over M_D^2 - M_{S_n^2}} ~ C_{S_n \gamma\gamma}
\ \ .
\label{mix0}
\eeqa
Let us consider two distinct kinds of contributions, 
$B_{D^0 \gamma\gamma}^{\rm mix} = 
B_{D^0 \gamma\gamma}^{\rm (gnd)} + B_{D^0 \gamma\gamma}^{\rm (res)}$:

\begin{enumerate}
\item If the spinless meson is a ground-state particle 
($\pi^0$, $\eta$ or $\eta'$),\footnote{The kaon intermediate 
state is disfavored due to the small $K\to \gamma \gamma$ 
branching ratio.} we have 
\beq
B_{D^0 \gamma\gamma}^{\rm gnd} 
= - {G_F a_2 f_D \alpha \over \sqrt{2} \pi} 
\left[ {\xi_d \over \sqrt{2}} {M_\pi^2 \over M_D^2 - M_\pi^2} 
+ {2 \xi_s - \xi_d \over 3 \sqrt{2}} \sum_{k = \eta,\eta'}
{M_k^2 \over M_D^2 - M_k^2} f_k (\theta) \right]
\ \ ,
\label{mix1}
\eeq
where $a_2 \simeq - 0.55$, $\theta \simeq -20^o$, 
$f_\eta (\theta) \equiv \cos^2 \theta 
- 2 \sqrt{2} \sin\theta \cos \theta$ and 
$f_{\eta'} (\theta) \equiv \sin^2 \theta 
+ 2 \sqrt{2} \sin\theta \cos \theta$.  The above parameterization 
for the two-photon vertices agrees with the values determined 
experimentally, 
\beq
B_{P_n \gamma\gamma} = \left\{ 
\begin{array}{ll}
0.0249~{\rm GeV}^{-1} & (\pi^0) \\
0.0275~{\rm GeV}^{-1} & (\eta) \\
0.0334~{\rm GeV}^{-1} & (\eta') \ \ .
\end{array} \right.
\label{mix2}
\eeq
$B_{D^0 \gamma\gamma}^{\rm gnd}$ is seen to vanish, 
as it must, in the limit of SU(3) flavor symmetry (there 
$\langle \eta' | {\cal H}_{wk}^{\rm (p.c.)} | D^0 \rangle = 0$ 
and the $\pi^0$, $\eta$ contributions cancel). From 
Eq.~(\ref{dgg2}), we obtain the branching ratio 
\begin{equation}
{\cal B}r_{D^0 \to\gamma\gamma}^{\rm gnd} \ \simeq \ 3 \times 
10^{-11} \ \ .
\label{mix3}
\end{equation}

\item If the intermediate meson is a spinless resonance $R^0$, 
the decay chain becomes $D^0 \to R^0 \to \gamma\gamma$.  
Since little is yet known about meson 
excitations, both the weak mixing amplitudes and the two-photon 
emission amplitudes must be modeled theoretically.  
The $D^0$-to-resonance weak matrix element 
will depend upon the flavor structure of $R^0$, {\it e.g.} 
\beq
\langle R^0 | {\cal H}_{wk}^{\rm (p.c.)} | D^0 \rangle 
= - {G_F a_2 f_D \over \sqrt{2}} \left\{ 
\begin{array}{ll}
\xi_d f_R /\sqrt{2} & (R^0 = ({\bar u}u - {\bar d}d)/\sqrt{2}) \\
\xi_s f_R & (R^0 = {\bar s}s) \\
V_{cd}^* V_{us} f_R & (R^0 = {\bar s}d)) \ \ ,
\end{array} \right.
\label{mix4}
\eeq
where the flavor content of $R^0$ is in parentheses and 
estimates for resonance decay constants $f_R$ are given in 
Ref.~\cite{GP98}.  The $R^0 \to \gamma\gamma$ mode has been 
observed for a number of resonances and has typical branching 
ratios ${\cal B}r_{R^0 \to\gamma\gamma} = {\cal O}(10^{-5})$ 
for $M_R \simeq 1 \to 1.3$~GeV, decreasing to 
${\cal B}r_{R^0 \to\gamma\gamma} = {\cal O}(10^{-6})$ 
for $M_R \ge 1.5$~GeV.  

For a concrete example of the resonance mechanism, 
we choose $R^0 = \pi(1800)$ and assume ${\cal B}r_{\pi(1800)\to
\gamma\gamma} \simeq 10^{-6}$.  The resulting $D^0 \to \gamma \gamma$ 
branching ratio is 
\begin{equation}
{\cal B}r_{D^0 \to\gamma\gamma}^{R^0 = \pi(1800)} \ \sim \ 
10^{-10}  \ \ .
\label{mix5}
\end{equation}
\end{enumerate}

\begin{center}
{\bf Two-particle Unitarity Contribution}  
\end{center}

\begin{figure}
\vskip .1cm
\hskip 2.5cm
\epsfig{figure=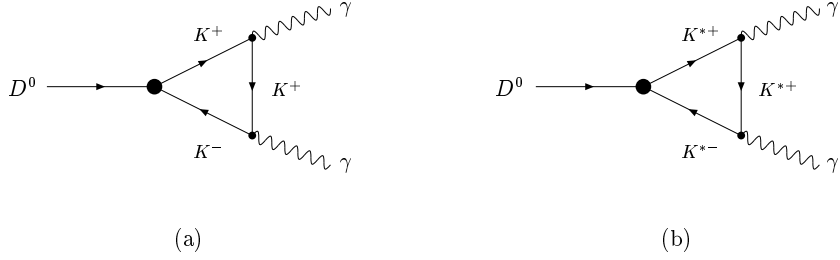,height=1.3in}
\caption{Unitarity contributions: (a) $K^+ K^-$, 
(b) $K^{*+} K^{*-}$.} 
\label{fig:dggfig3}
\end{figure}

In a factorization approach, the $D^0 \to K^+ K^-$ amplitude ({\it cf.}
Fig.~\ref{fig:dggfig3}) is 
\beq
{\cal M}_{D^0 K^+ K^-} = {G_F M_D^2 \over \sqrt{2}}V_{cs} V_{us}^* 
f \left[ \left( 1 - {M_K^2 \over M_D^2} \right) f_+ (M_K^2) + 
{M_K^2 \over M_D^2} f_- (M_K^2) \right] \ \ ,
\label{u1}
\eeq
where $f_\pm$ are form factors and $f$ is a constant 
containing information about QCD corrections and the kaon decay 
constant.  A fit to the measured $D^0 \to K^+ K^-$ decay rate yields 
\beq
f \left[ \left( 1 - {M_K^2 \over M_D^2} \right) f_+ (M_K^2) + 
{M_K^2 \over M_D^2} f_- (M_K^2) \right] = 141~{\rm MeV} 
\ \ .
\label{u2}
\eeq
Similar to the $B_s$ system\cite{ellisch},
the $K^+ K^-$ intermediate state contributes via unitarity 
to only the amplitude $C_{D^0 \gamma\gamma}$ of Eq.~(\ref{dgg1}) and 
is proportional to precisely the same combination of form factors 
appearing in Eq.~(\ref{u2}), 
\beq
{\cal I}m~C_{D^0 \gamma\gamma}^{(K^+ K^-)} = 2 \alpha {M_K^2 \over M_D^4} 
\sqrt{ 1 - 4 M_K^2 / M_D^2} ~{\cal M}_{D^0 K^+ K^-} \ \ ,
\label{u3}
\eeq
from which we obtain 
\beq
{\cal B}r_{D^0 \to\gamma\gamma}^{(K^+ K^-)} \sim 0.7 \times 
10^{-8} \ \ .
\label{u5}
\eeq

\begin{center}
{\bf Summary of $D^0 \to \gamma\gamma$}
\end{center}

Considered together, the above examples lead us to 
anticipate a branching ratio in the neighborhood of 
$10^{-8}$.  Our maximal ({\it i.e.} constructive interference) 
VMD signal has a central value ${\cal B}r_{D^0 \to\gamma 
\gamma}^{\footnotesize{\rm (vmd)}} \simeq 3.5 \times 10^{-8}$.  
The recent work of Ref.~\cite{fsz} 
provides an independent estimate of the $D^0 \to \gamma\gamma$ 
transition and obtains a similar order-of-magnitude result.

\subsection{Lepton-antilepton Emission $D^0\to \ell^+ \ell^-$}

The general form for the amplitude describing 
$D^0 (p) \to \ell^+(k_+,s_+) \ell^-(k_-,s_-)$ is 
\begin{equation}
{\cal M}_{D^0 \to \ell^+ \ell^-} = {\bar u}(k_-,s_-) 
\left[A_{D^0\ell^+\ell^-} \ + \ \gamma_5 ~B_{D^0\ell^+\ell^-} 
 \right] v(k_+ , s_+) \ \ ,
\label{2ca}
\end{equation}
and the associated decay rate is 
\begin{equation}
\Gamma_{D^0 \to \ell^+ \ell^-} = {M_D \over 8 \pi} 
\sqrt{1 - 4 {m_\ell^2 \over M_D^2}}  
\left[ |A_{D^0\ell^+\ell^-} |^2 
\ + \ \left( 1 - 4 {m_\ell^2 \over M_D^2}\right) 
|B_{D^0\ell^+\ell^-} |^2 \right] \ \ .
\label{2cb}
\end{equation}

\subsubsection{\bf Short Distance Contributions 
$c {\bar u} \to \ell^+ \ell^-$}

The short distance (${\cal O}(\alpha_s)$ corrected) transition 
amplitude is given by~\cite{buras}
\begin{equation}
B_{D^0\ell^+\ell^-}^{\rm (s.d.)} \simeq 
{G_F^2M_W^2f_Dm_\ell \over \pi^2} F \,,
\label{d2musm}
\end{equation}
where
\begin{equation}
F=\sum_{i=d,s,b} \ V_{ui}V^*_{ci}\left[ {x_i \over 2}
+{\alpha_s \over 4 \pi}x_i \cdot \left( \ln^2 x_i 
+ {4 + \pi^2 \over 3}\right) \right] \,,
\end{equation}
with $x_i=m_i^2/M_W^2$.  The amplitude $A_{D^0\ell^+\ell^-}$ vanishes
due to the equations of motion.  The explicit dependence on lepton mass 
in the decay amplitude overwhelmingly favors the $\mu^+\mu^-$ 
final state over that of $e^+e^-$. 
Upon employing the quark mass values $m_d\simeq 0.01$~GeV, 
$m_s\simeq 0.12$~GeV, $m_b\simeq 5.1$~GeV, the Wolfenstein 
CKM parameters $\lambda \simeq 0.22$, $A \simeq 0.82$, 
$\rho \simeq 0.21$, $\eta \simeq 0.35$ and the decay constant 
$f_D\simeq 0.2$~GeV, we obtain the branching fraction 
${\cal B}r^{s.d.}_{D^0 \to\mu^+\mu^-} \simeq 10^{-18}$.

\subsubsection{Long Distance Contributions to $D^0 \to \ell^+ \ell^-$}

\begin{figure}
\vskip .1cm
\hskip 1.5cm
\epsfig{figure=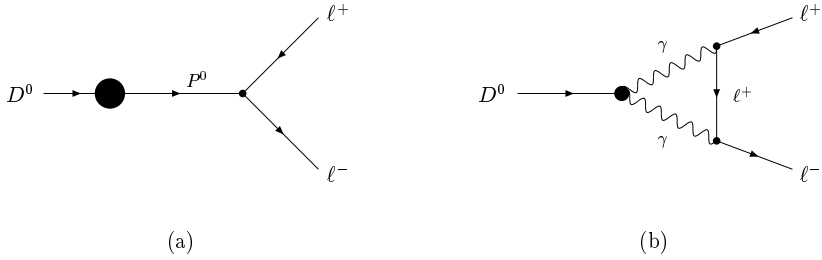,height=1.3in}
\caption{Unitarity contributions: (a) One-particle, 
(b) Two-particle $\gamma\gamma$.}
\label{fig:dggfig5}
\end{figure}

In the following, we consider two 
long distance unitarity contributions ({\it cf.} Fig.~\ref{fig:dggfig5})
which lead to $D^0 \to \ell^+ \ell^-$ transitions.  In each case, the 
decay amplitude is dependent on the lepton mass, and thus 
we shall provide numerical branching ratios only for the case 
$D^0 \to \mu^+ \mu^-$. 

\begin{center}
{\bf Single-particle Unitarity Contribution}  
\end{center}

The single-particle `weak-mixing' contribution to $D^0\to\ell^+\ell^-$ can be 
estimated in a manner like that considered for the 
$D^0 \to \gamma \gamma$ transition ({\it cf.} Eq.~(\ref{mix0})).   
For definiteness, we consider the $D^0\to\ell^+\ell^-$ parity-conserving 
amplitude $B_{D^0\ell^+\ell^-}$ (see Eq.~(\ref{2ca})),  
\beq
B_{D^0\ell^+\ell^-}^{\rm (mix)}\  = \ \sum_{P_n} \ 
\langle P_n | {\cal H}_{wk}^{\rm (p.c.)} | D^0 \rangle 
~{1 \over M_D^2 - M_{P_n^2}} ~B_{P_n \ell^+ \ell^-}\ \ ,
\label{ellmix0}
\eeq
and we write 
$B_{D^0\ell^+\ell^-}^{\rm (mix)} = B_{D^0\ell^+\ell^-}^{\rm (gnd)} 
\ + \ B_{D^0\ell^+\ell^-}^{\rm (res)}$ 
for the ground state ($\pi^0,\eta,\eta'$) and resonance contributions.

There is little known regarding the 
$P_n \mu^+ \mu^-$ ($P_n = \pi^0,\eta,\eta'$) vertices.  
In the following, we assume these quantities have 
the same flavor structure as the corresponding 
$P_n \gamma \gamma$ vertices described 
earlier,\footnote{This ensures that our expression 
will vanish in the limit of SU(3) flavor symmetry.}  
and obtain the overall $P_n \mu^+ \mu^-$ 
normalization from the measured 
$\eta \to \mu^+ \mu^-$ mode.  From this we predict 
for the $\eta'~(960) \to \mu^+ \mu^- $ mode a branching ratio 
${\cal B}r_{\eta' \mu^+ \mu^-} \simeq 5.6 \times 
10^{-7}$, well below the current bound 
${\cal B}r_{\eta' \mu^+ \mu^-} < 10^{-4}$.
The ground state contribution is then 
\beqa
B^{\rm (gnd)}_{D^0\ell^+\ell^-} &=& - {G_F a_2 f_D 
B_{P \mu^+\mu^-} \over \sqrt{2}} 
\bigg[ {\xi_d \over \sqrt{2}} 
{M_\pi^2 \over M_D^2 - M_\pi^2} 
\nonumber \\
& & + {2 \xi_s - \xi_d  \over 3 \sqrt{2}} 
{M_\eta^2 \over M_D^2 - M_\eta^2} \left( \cos^2 \theta 
- 2 \sqrt{2} \sin\theta \cos \theta \right) 
\nonumber \\
& & + {2 \xi_s - \xi_d  \over 3 \sqrt{2}} 
{M_\eta^{'2} \over M_D^2 - M_\eta^{'2}} \left( \sin^2 \theta 
+ 2 \sqrt{2} \sin\theta \cos \theta \right) \bigg] \ \ ,
\label{ellmix3}
\eeqa
with $B_{P \mu^+\mu^-} = 3.47 \times 10^{-5}$.  This leads
to the branching ratio 
\begin{equation}
{\cal B}r_{D^0\to\ell^+\ell^-}^{\rm (gnd)} \ \simeq \ 2.5 \times 
10^{-18} \ \ .
\label{ellmix4}
\end{equation}

There can also, in principle, be intermediate state contributions 
from $J^P = 0^\pm$ neutral resonances $\{ R^0 \}$.  Using 
the $D^0$-to-$R^0$ mixing amplitude already obtained in 
Eq.~(\ref{mix4}) and again identifying the resonance $R^0$ as 
$\pi(1800)$, we find 
\beq
{\cal B}r_{D^0\to \ell^+\ell^-}^{\rm (\pi (1800))} \ \simeq \ 
1.8 \times 10^{-3} {\Gamma_{\pi (1800) \ell^+\ell^-} \over M_{\pi(1800)}} 
\ = \ 1.8 \times 10^{-3} {\cal B}r_{\pi(1800)\to\ell^+\ell^-}
\label{ellmix5}
\eeq
Upon assuming ${\cal B}r_{\pi(1800)\to\ell^+\ell^-} = 10^{-12}$ 
as our default branching ratio, we obtain 
\beq
{\cal B}r_{D^0\to \ell^+\ell^-}^{\rm (\pi(1800))} \ \simeq \ 
5.0 \times 10^{-17} {{\cal B}r_{\pi(1800)\to \ell^+\ell^-} 
\over 10^{-12}} \ \ .
\label{ellmix6}
\eeq
Although possibly enhanced relative to the light-meson pole 
contributions, the result is still unmeasureably small.

\begin{center}
{\bf The Two-photon Unitarity Contribution}
\end{center}

In the $K_L \to e^+ e^-$ transition, the 
two-photon intermediate state is known to 
play an important role.  Let us therefore consider the 
contribution of this intermediate state for $D^0 \to \ell^+ \ell^- $, 
\begin{eqnarray}
{\cal I}m~ {\cal M}_{D^0 \to \ell^+ \ell^-} &=& 
{1 \over 2!} \sum_{\lambda_1,\lambda_2} \int 
{d^3 q_1 \over 2\omega_1 (2 \pi)^3} 
~{d^3 q_2 \over 2\omega_2 (2 \pi)^3}
\\ 
& & \times\ {\cal M}_{D \to \gamma \gamma} \ {\cal M}^*_{\gamma \gamma \to 
\ell^+ \ell^-} (2 \pi)^4 \delta^{(4)} (p - q_1 - q_2) \ \ .
\label{2gb}
\end{eqnarray}
Upon inserting the general form of the $D^0 \to \gamma\gamma$ 
appearing in Eq.(\ref{dgg3}), we obtain 
\begin{eqnarray}
{\cal I}m ~A_{D^0 \ell^+ \ell^-}^{(\gamma\gamma)} =  
\alpha m_\ell B_{D^0 \gamma \gamma} \ln {M_D^2 \over m_\ell^2}
\ , \qquad 
{\cal I}m~ B_{D^0 \ell^+ \ell^-}^{(\gamma\gamma)} =
i \alpha m_\ell C_{D^0 \gamma \gamma} \ln {M_D^2 \over m_\ell^2}\ .
\label{2gc}
\end{eqnarray}
We find
\begin{equation}
{\cal B}r_{D^0\to \mu^+\mu^-}^{\rm (\gamma\gamma)} \ 
\simeq \ 2.7 \times 10^{-5} 
{\cal B}r_{D^0 \to \gamma \gamma} \ \ .
\label{2gd}
\end{equation}

\begin{center}
{\bf Summary of $D^0 \to \mu^+\mu^-$}
\end{center}

The largest of our estimates, the two-photon unitarity 
component, for the long distance contribution to 
$D^0 \to \mu^+\mu^-$ favors a branching ratio somewhere 
in excess of $10^{-13}$.  More generally, it scales as 
$2.7 \times 10^{-5}$ times the branching ratio for  
$D^0 \to \gamma \gamma$.  With the estimate 
${\cal B}r_{D^0 \to\gamma \gamma} \ge  10^{-8}$ 
arrived at in the previous section, we therefore anticipate 
a branching ratio for $D^0 \to \mu^+\mu^-$ of at least 
$3 \times 10^{-13}$.

\section{Potential for New Physics Contributions}
\label{sec:bsm}

As discussed in the introduction, the charm system provides a unique
laboratory to probe physics beyond the Standard Model as it offers
a complementary probe of physics to that attainable from a study of
rare processes in the down-quark sector.  As we found in the previous
section, short distance SM contributions to rare charm
decays are quite small due to the effectiveness
of the GIM mechanism, and most reactions are dominated by long range
effects.  However, we saw that for some reactions there exists a window for
the potential observation of new short distance effects, in 
particular for specific
regions of the invariant dilepton mass spectrum in $D\to X\ell^+\ell^-$.
Indeed in some cases, it is precisely because the SM rates are so small
that charm provides an untapped opportunity to discover new effects and 
offers a detailed test of the SM in the up-quark sector.

In this section, we delineate some new physics possibilities, motivated
by supersymmetric, grand-unified, extra dimensional, or strongly 
coupled extensions of the SM, which give rise to observable effects
in rare charm transitions.  In some cases, we find that present 
experimental limits on these channels already constrain the model
parameter space.

\subsection{\bf Supersymmetry and Rare Charm Decays}
\label{sec:susy}

We first examine the effects of Supersymmetry (SUSY) in rare charm decays,
concentrating on the exclusive modes $D\to \pi\ell^+\ell^-$ and
$D\to\rho\ell^+\ell^-$.  Weak scale Supersymmetry is a possible solution 
to the hierarchy problem and as such is a well motivated theory of
physics beyond the SM.  We consider the general case of the
unconstrained version of the Minimal Supersymmetric extension of
the Standard Model where no particular SUSY breaking mechanism
is assumed and investigate the two scenarios where R-parity is
conserved or violated.  Imposing the constraints on the SUSY
parameter space from current data, we find that in both cases, the 
supersymmetric contributions to these decay channels can be quite large, 
particularly in the low dilepton mass region ({\it i.e.} below $m_\rho$).

\subsubsection{\bf Minimal Supersymmetric Standard Model}
\label{sec:mssm}

\begin{figure}[t]
\vskip-4cm\hspace*{3.0cm}
\epsfig{figure=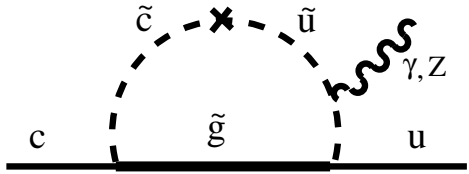,height=4.2in,angle=-90}
\vskip-2.5cm
\caption{A typical contribution to $c\to u$ FCNC transitions in 
the MSSM. The cross denotes one  mass insertion 
$(\delta^u_{12})_{\lambda\lambda'}$, with $\lambda,\lambda'=L,R$.}
\label{mssm}
\end{figure}

The Minimal Supersymmetric Standard Model (MSSM) is the simplest 
supersymmetric extension of the SM and involves a doubling of the 
particle spectrum by putting all SM fermions in chiral supermultiplets, 
as well as the SM gauge bosons in vector supermultiplets. 
In our discussion,  we do not assume any particular Supersymmetry 
breaking mechanism, but rather use a parameterization of all possible soft SUSY
breaking terms.  A large number, of order 100, 
of new parameters is then introduced. The soft supersymmetry 
breaking sector generally includes three gaugino masses, as well as trilinear 
scalar interactions, Higgs and sfermion masses. 
Supersymmetry contains many potential sources for flavor violation.  In
particular, if
we choose to rotate the squark fields by the same matrices that 
diagonalize the quark mass matrices, then the squark mass matrices 
are not diagonal.  
In this super-CKM basis, squark propagators can be expanded 
so that non-diagonal mass terms result in mass insertions that change 
the squark flavor~\cite{lhall,susyfcnc}.   These mass insertions can be 
parameterized in a model independent fashion via
\begin{equation}
(\delta^u_{ij})_{\lambda\lambda'}={(M^u_{ij})^2_{\lambda\lambda'}
\over M^2_{\tilde q}}\,,
\end{equation}
where $i\neq j$ are generation indices, $\lambda,\lambda'$ denote the
chirality, $(M^u_{ij})^2$ are the off-diagonal elements of the up-type squark 
mass matrix, and $M_{\tilde q}$ represents the average squark mass.
The exchange of squarks in loops thus leads to FCNC
through diagrams such as the one depicted in Fig.~\ref{mssm}.  This 
source of flavor violation
can be avoided in specific SUSY breaking scenarios such as gauge-mediation
or anomaly mediation, but is present in general. It appears, for instance
if SUSY breaking is mediated by gravity.

The MSSM contributions to loop mediated processes in addition to those
of the SM  are: gluino-squark exchange, chargino/neutralino-squark
exchange and charged Higgs-quark exchange. This last contribution
carries the same CKM structure as in the SM
loop diagram and is proportional to the internal and external quark
masses; it thus leads to small effects in rare charm transitions and
we neglect it here.  The gluino-squark contribution proceeds via flavor
diagonal vertices proportional to the strong coupling constant and in
principle dominates the CKM suppressed, weak-scale strength
chargino/neutralino-squark contributions.  We thus only consider the
case of gluino-squark exchange here as an estimate of the potential size
of supersymmetric effects in rare charm decays.  
We note that the analogous gluino
contributions to rare $K$ and $B$ transitions have led to
strong universality constraints on the charged $Q=-1/3$ squark
sector~\cite{ellis}.  Here, we examine the level at which the corresponding
constraints can be obtained in the charged $Q=+2/3$ squark sector once
data accumulates at $B$ and charm factories.
 
Within the context of the mass insertion approximation the effects
are included in the Wilson 
coefficients corresponding to the decay $D\to X\ell^+\ell^-$ via
\begin{equation}
C_i = C_i^{SM} + C_i^{\tilde g}\,,
\end{equation}
for $i=7,9,10$.
Allowing for only one insertion, the explicit contributions 
from the gluino-squark diagrams are~\cite{lunghi,wyler}
\begin{equation}
C_7^{\tilde g} = -\frac{8}{9}\,\frac{\sqrt 2}{G_FM_{\tilde q}^2}
\pi\alpha_s\,\left\{ (\delta^u_{12})_{LL}\frac{P_{132}(u)}{4} 
 + (\delta^u_{12})_{LR}P_{122}(u)\frac{M_{\tilde g}}{m_c}
\right\}~,
\label{c7g}
\end{equation}
and 
\begin{equation}
C_9^{\tilde g} = -\frac{8}{27}\,\frac{\sqrt 2}{G_FM_{\tilde q}^2}
\pi\alpha_s\,(\delta^u_{12})_{LL}P_{042}(u)~,
\label{c9g}
\end{equation}
with the contribution to $C_{10}$ vanishing at this order due to the
helicity structure.
If we allow for two mass insertions, there is a contribution to 
$C_{9,10}$ given by
\begin{equation}
C_{10}^{\tilde g} = -\frac{1}{9}\,\frac{\alpha_s}{\alpha}\,
(\delta^u_{22})_{LR}  (\delta^u_{12})_{LR} P_{032}(u) = -{C_9\over
1-4\sin^2\theta_W}~.
\label{c10g}
\end{equation}
Here, 
$u=M_{\tilde g}^2/M_{\tilde q}^2$
and the functions
$P_{ijk}(u)$ are defined as 
\begin{equation}
P_{ijk}(u) \equiv \int_{0}^{1} dx \ \frac{x^i(1-x)^j}{(1-x+ux)^k}~.
\label{pdef} 
\end{equation} 
In addition, the operator basis can be extended by the ``wrong chirality'' 
operators $\hat O_7$, $\hat O_9$ and $\hat O_{10}$, obtained  by switching the 
quark chiralities in Eqs.~(\ref{gdipole}) and (\ref{laxial}). The 
gluino-squark contributions to the corresponding Wilson coefficients
are
\begin{eqnarray}
\hat C_7^{\tilde g} &=& -\frac{8}{9}\,\frac{\sqrt 2}{G_FM_{\tilde q}^2}
\pi\alpha_s\,\left\{ (\delta^u_{12})_{RR}\frac{P_{132}(u)}{4} 
 + (\delta^u_{12})_{LR}P_{122}(u)\frac{M_{\tilde g}}{m_c}
\right\}~,
\label{c7pg}\\
\hat C_9^{\tilde g} &=& -\frac{8}{27}\,\frac{\sqrt 2}{G_F M_{\tilde q}^2}
\pi\alpha_s\,(\delta^u_{12})_{RR}P_{042}(u)-(1-4\sin^2\theta_W)\hat 
C^{\tilde g}_{10}~,\nonumber\\
\hat C_{10}^{\tilde g} &=& -\frac{1}{9}\,\frac{\alpha_s}{\alpha}\,
(\delta^u_{22})_{LR}  (\delta^u_{12})_{LR} P_{032}(u) \ \ , \nonumber
\end{eqnarray}
where the expression for $\hat C_{10}^{\tilde g}$ is again obtained with 
a double insertion. 

As was noted in Refs.~\cite{lunghi,wyler}, in both $C_7^{\tilde g}$ 
and $\hat C_7^{\tilde g}$ the term in which the squark chirality labels 
are mixed introduces the enhancement factor 
$M_{\tilde g}/m_c$. In the SM the chirality flip which appears in 
$O_7$ occurs by a 
flip of one external quark line, resulting in
a factor of $m_c$
included in the operator's definition\footnote{ The 
$m_u$ term, proportional to the ($1-\gamma_5$) in the operator, is neglected.}.
However, in the gluino-squark diagram, the insertion of 
$(\delta^u_{12})_{RL}$ forces the chirality flip to take place 
in the gluino line, thus introducing a $M_{\tilde g}$ factor instead of
$m_c$.  This yields a significant enhancement in the short
distance contributions to the process
$D\to X_u\gamma$~\cite{wyler}, which is unfortunately obscured by the 
large long range effects.

The most stringent bounds that apply to the non-universal soft breaking
terms $(\delta^u_{12})_{\lambda\lambda'}$ come from the experimental 
searches for $D^0-\bar{D^0}$ mixing\footnote{Limits obtained from 
charge and color breaking (CCB) and bounding the potential from 
below (UFB)~\cite{savas} apply to the trilinear terms but not to the 
squark mass terms. Thus, unless the squark mass  matrices are kept diagonal, 
CCB and UFB arguments cannot be used to constrain the non-universal mass
insertions.}. The current CLEO limit~\cite{cleodd} 
implies~\cite{wyler} 
\begin{equation}
\frac{1}{2}\left\{\left(\frac{\Delta m_D}{\Gamma_{D^0}}\right)^2\cos\delta
+\left(\frac{\Delta\Gamma_D}{2\Gamma_{D^0}}\right)^2\sin\delta\right\}
<0.04\%~,
\label{ddbound}
\end{equation}  
where $\delta$ is a strong relative phase between the Cabibbo-allowed and 
the doubly Cabibbo-suppressed $D^0\to K\pi$ decays. Neglecting this phase 
results in the constraints obtained in Ref.~\cite{wyler}, 
which we collect in Table~\ref{numssm}. 
\begin{table}[h]
\centering
\begin{tabular}{|c|c|c|} \hline\hline
$M^2_{\tilde g}/M^2_{\tilde q}$ & $(\delta^u_{12})_{LL}$ & 
$(\delta^u_{12})_{LR}$ 
\\ \hline
$0.3$ & $0.03$ & $0.04$ \\
$1.0$ & $0.06$ & $0.02$ \\
$4.0$ & $0.14$ & $0.02$ \\
\hline\hline
\end{tabular}
\caption
{Bounds on $(\delta^u_{12})_{LL}$, $(\delta^u_{12})_{LR}$ from 
$D^0-\bar{D^0}$
mixing~\cite{wyler} (neglecting the strong phase). All constraints should be 
multiplied by $(M_{\tilde q}/500{~\rm GeV})$. 
}
\label{numssm}
\end{table}
These bounds were obtained assuming 
that $(\delta^u_{12})_{RR}=0$ and $(\delta^u_{12})_{LR}=(\delta^u_{12})_{RL}$;
these assumptions are found to be numerically unimportant.

\begin{figure}
\hskip 3.0cm
\epsfig{figure=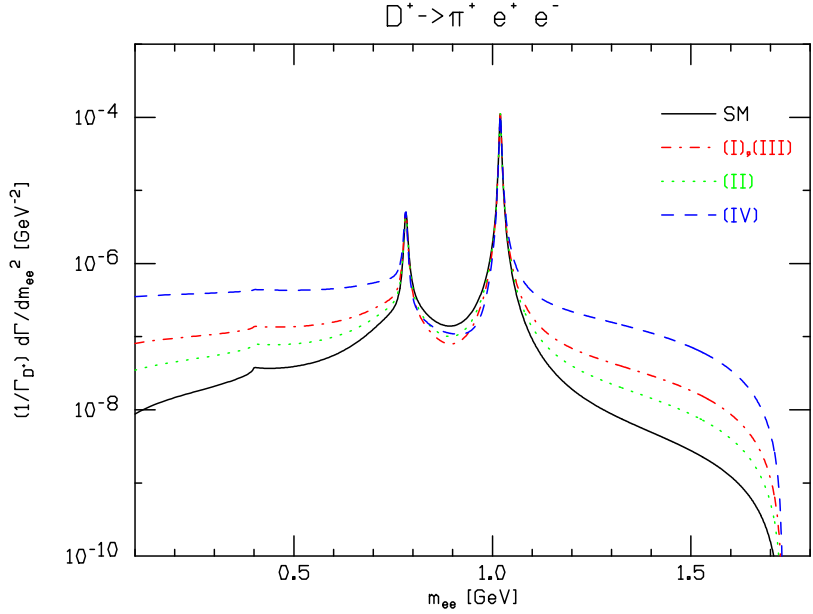,height=4.2in,angle=90}
\caption{The dilepton mass distribution for  
$D^+\to\pi^+ e^+e^-$ (normalized to $\Gamma_{D^+}$), 
in the MSSM with non-universal 
soft breaking effects.
The solid line is the SM; (I): $M_{\tilde g}=M_{\tilde q}=250$~GeV; 
(II): $M_{\tilde g}=2\,M_{\tilde q}=500$~GeV; 
(III): $M_{\tilde g}=M_{\tilde q}=1000$~GeV and 
(IV): $M_{\tilde g}=(1/2)\,M_{\tilde q}=250$~GeV.
}
\label{pill_mssm}
\end{figure}

\begin{figure}
\hskip 3.0cm
\epsfig{figure=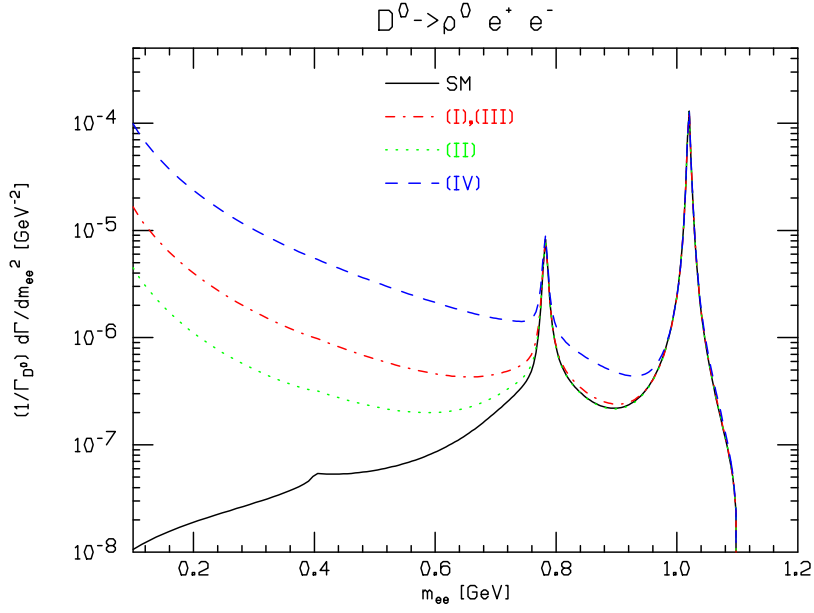,height=4.2in,angle=90}
\caption{The dilepton mass distribution for  
$D^0\to\rho^0 e^+e^-$ (normalized to $\Gamma_{D^0}$),
in the MSSM with non-universal 
soft breaking effects.
The solid line is the SM; (I): $M_{\tilde g}=M_{\tilde q}=250$~GeV; 
(II): $M_{\tilde g}=2\,M_{\tilde q}=500$~GeV; 
(III): $M_{\tilde g}=M_{\tilde q}=1000$~GeV and 
(IV): $M_{\tilde g}=(1/2)\,M_{\tilde q}=250$~GeV.
}
\label{rholl_mssm}
\end{figure}

In order to estimate the effects in $c\to u\ell^+\ell^-$ transitions from
the gluino contributions,
we need to specify $M_{\tilde g}$ and $M_{\tilde q}$. We consider 
four sample cases: (I): $M_{\tilde g}=M_{\tilde q}=250$~GeV; 
(II): $M_{\tilde g}=2\,M_{\tilde q}=500$~GeV; 
(III): $M_{\tilde g}= M_{\tilde q}=1000$~GeV and 
(IV): $M_{\tilde g}=(1/2)\,M_{\tilde q}=250$~GeV.
We first examine $D^+\to\pi^+e^+e^-$. In Fig.~\ref{pill_mssm}
we show the dilepton mass distribution as a function of the dilepton mass.
Although the net effect is relatively small in the integrated rate (an increase
$\simeq 20\%$
or smaller), the enhancement due to the SUSY contributions is most conspicuous
away from the vector resonances, particularly for low dilepton masses. 
Experiments sensitive to the dilepton mass distribution 
at the level of $10^{-7}-10^{-8}$ can detect these SUSY contributions.  
However, the decays to a  vector meson, such  as 
$D\to\rho e^+ e^-$, are more sensitive to the gluino exchange, 
as can be seen from Fig.~\ref{rholl_mssm}. 
The effect is quite pronounced and almost entirely
lies in the low $m_{ee}$ region. This is mostly
due to the contributions of 
$(\delta^u_{12})_{RL}$ to $C_7$ and $\hat C_7$ in Eqs.~(\ref{c7g}) 
and (\ref{c7pg}), which contain the $M_{\tilde g}/m_c$ enhancement as discussed
above. This effect is intensified at low $q^2=m_{ee}^2$ due to the 
photon propagator (see for instance Eq.~(\ref{dbs}) for the inclusive decays). 
This low $q^2$ enhancement of the $O_7$ contribution is present in exclusive
modes with vector mesons such as $D\to\rho\ell^+\ell^-$, but not in modes
with pseudoscalars, such as $D\to\pi\ell^+\ell^-$, since gauge invariance
forces a cancellation of the $1/q^2$ factor in the latter case ({\it e.g.,} 
see Eq.~(\ref{meedist})).
This is   apparent from a comparison of the low dilepton mass regions
in Figs.~(\ref{pill_mssm}) and~(\ref{rholl_mssm}). 

We conclude that the 
$D\to\rho\ell^+\ell^-$ decays are considerably sensitive to non-universal
soft breaking in the MSSM. 
The largest effect is obtained in case (IV) (dashed line in 
Fig.~(\ref{rholl_mssm})) and yields ${\cal B}r_{D^0\to\rho^0 e^+ e^-}
\simeq 1.3\times 10^{-5}$, which is roughly a factor of five times larger than 
the SM prediction given in Sect.~\ref{longd_sec}. The current 
experimental bound on this channel is~\cite{e791_1} 
${\cal B}r^{\rm exp}_{D^0\to\rho^0e^+e^-}<1.2\times10^{-4}$.  For muon
final states, the somewhat more stringent constraint
${\cal B}r^{\rm exp}_{D^0\to\rho^0\mu^+\mu^-}<2.2\times10^{-5}$ 
should be compared to ${\cal B}r_{D^0\to\rho^0 \mu^+ 
\mu^-}\simeq 1.3\times 10^{-6}$ obtained in 
case (IV). Thus, searches for rare charm decays with sensitivities of 
$10^{-6}$ and better will soon constrain the MSSM parameter space or
observe an effect.

\subsubsection{\bf R Parity Violation} 
\label{rpvsection}

The assumption of $R$-parity conservation in the MSSM  prohibits 
baryon and lepton number violating terms in the 
super-potential.  However, other symmetries can be invoked to prohibit rapid 
proton decay, such as baryon-parity or lepton-parity~\cite{ross}, and hence
allow for R parity
violation. The $R$-parity violating super-potential 
can be written as\footnote{We ignore bilinear terms which are 
not relevant to our discussion of FCNC effects.} 
\begin{equation}
{\cal W}_{R_p} =\epsilon_{ab}\left\{ 
\frac{1}{2}\lambda_{ijk}L^a_iL^b_j\bar{E}_k
+\lambda'_{ijk}L_i^aQ^b_j\bar{D}_k 
+\frac{1}{2}\epsilon_{\alpha\beta\gamma}\lambda^{''}_{ijk}\bar{U}^\alpha_i
\bar{D}^\beta_j\bar{D}^\gamma_k \right\}~,
\label{rpv}
\end{equation}
where $L$, $Q$, $\bar E$, $\bar U$ and $\bar D$ are the chiral super-fields
in the MSSM. The $SU(3)$ color indices are denoted by 
$\alpha,\beta,\gamma=1,2,3$, the 
$SU(2)_L$ indices by $a,b=1,2$ and the generation indices are $i,j,k=1,2,3$. 
The fields in Eq.~(\ref{rpv}) are in the weak basis.  The $\lambda'_{ijk}$
term is the one which is relevant for the rare charm decays we consider here
as it can give rise to tree-level contributions through the exchange of
squarks to decay channels such as $D\to X\ell^+\ell^-$, 
$D\to\ell^+\ell^-$, 
as well as the lepton-flavor violating $D\to X\mu^+e^-$ and 
$D\to\mu^+ e^-$ modes.  Before considering the FCNC effects in $D$ 
decays, we need to rotate the fields to the mass basis.
This leads to 
\begin{equation}
{\cal W}_{R_p}= \tilde{\lambda'}_{ijk}\left[N_i V_{jl} D_l
-E_iU_j\right] \bar{D}_k+\cdots~
\label{mbasis}
\end{equation}
where $V$ is the CKM matrix and we define
\begin{equation}
\tilde{\lambda'}_{ijk}\equiv \lambda'_{irs} {\cal U}^L_{rj} 
{\cal D}^{*R}_{sk}~.
\label{newlambda}
\end{equation}
Here, ${\cal U}^L$ and ${\cal D}^R$ are the matrices used to rotate the 
left-handed up- and right-handed down-quark fields to the mass basis. 
Written in terms of component fields, this interaction now reads
\begin{eqnarray}
{\cal W}_{\lambda'}&=&\tilde{\lambda}'_{ijk} \left\{V_{jl}[ 
\tilde{\nu}_L^i\bar{d}_R^kd_L^l + \tilde{d}_L^l\bar{d}_R^k\nu_L^i 
+ (\tilde{d}_R^k)^*(\bar{\nu_L^i})^cd_L^l]\right.\nonumber\\
& &\left.
-\tilde{e}_L^i\bar{d}_R^ku_L^j - \tilde{u}_L^j\bar{d}_R^ke_L^i 
-(\tilde{d}_R^k)^*(\bar{e_L^i})^cu_L^j \right\}~.
\label{incomps}
\end{eqnarray}
The last term in Eq.~(\ref{incomps}) can give rise to the processes
$c\to u\ell\ell^{(')}$ at tree level via the exchange of a down-squark.
This leads to effects that are proportional to 
$\tilde{\lambda}'_{i2k}\tilde{\lambda}'_{i1k}$ with $i=1,2$ (due to
kinematical restrictions).
 
Constraints on these coefficients have been derived in the 
literature~\cite{rpv1}.
For instance, tight bounds are obtained in Ref.~\cite{agashe}
from $K^+\to\pi^+\nu\bar\nu$ by assuming that only one R-parity violating 
coupling satisfies $\tilde{\lambda}'_{ijk}\neq 0$. We update this bound 
by using the latest experimental result~\cite{k2pinunubar} 
${\cal B}r_{K^+ \to\pi^+\nu\bar\nu}=(1.57^{+ 1.75}_{- 0.82})\times10^{-10}$, 
which yields
$\tilde{\lambda}'_{ijk} <0.005$.  However, this bound can be avoided
in the single coupling scheme~\cite{agashe}, where only one R-parity
violating coupling is taken to be non-zero in the weak basis. 
In this case, it is possible that flavor rotations may restrict the
R-parity breaking induced flavor violation to be present in either the
charge $-1/3$ or $+2/3$ quark sectors, but not both.
Then large effects are possible in the up sector 
for observables such as $D^0$-$\bar{D^0}$ mixing and rare decays without
affecting the down-quark sector.
In Ref.~\cite{agashe}
a rather loose constraint on the R-parity breaking couplings is obtained from 
$D^0$ mixing, which could result in large effects in $c\to u\ell\ell^{(')}$ 
decays. Here, we will take a conservative approach and make use of more 
model-independent bounds. The constraints on the R-parity breaking couplings
for the processes of interest here
are collected in Table~\ref{rpvbounds} from Ref.~\cite{rpv1}.
The charged current universality bounds assume three
generations. The $\pi$ decay constraint is given by the quantity
$R_\pi=\Gamma_{\pi\to e\nu}/\Gamma_{\pi\to\mu\nu}$. The limits obtained from
$D\to K\ell\nu$ were first obtained in Ref.~\cite{rpv2}. 

\begin{table}[h]
\centering
\begin{tabular}{|c|c|c|c|} \hline\hline
 & & & \\
$\tilde{\lambda}'_{11k}$ & $\tilde{\lambda}'_{12k}$ & 
$\tilde{\lambda}'_{21k}$ & $\tilde{\lambda}'_{22k}$\\ \hline
$0.02^{\rm (a)}$ & $0.04^{\rm (a)}$ & $0.06^{\rm (b)}$ & $0.21^{\rm (c)}$ \\
\hline\hline
\end{tabular}
\caption
{ Most stringent ($2\sigma$) bounds for the $R$-parity violation couplings 
entering 
in rare $D$ decays, from (a) charged current universality,
(b) $R_\pi$ and (c) $D\to K\ell\nu$. See Ref.~\cite{rpv1}
for details. All numbers should be multiplied by 
$(m_{\tilde{d}^k_R}/100{\rm ~GeV)}$.}
\label{rpvbounds}
\end{table}

We first consider the contributions to $c\to u\ell^+\ell^-$. 
The tree level exchange of down squarks results in 
the effective interaction
\begin{equation}
\delta {\cal H}_{\rm eff} = -\frac{\tilde{\lambda}'_{i2k}\tilde{\lambda}'_{i1k}}
{m^2_{\tilde{d}^k_R}}\, \overline{(\ell_L)^c}c_L\,\,\bar{u}_L(\ell_L)^c~,
\label{dheff1}
\end{equation}  
which after Fierzing gives
\begin{equation}
\delta {\cal H}_{\rm eff} =
-\frac{\tilde{\lambda}'_{i2k}\tilde{\lambda}'_{i1k}}
{2 m^2_{\tilde{d}^k_R}}\,(\bar u_L\gamma_\mu c_L)(\bar\ell_L\gamma^\mu\ell_L)~.
\label{dheff2}
\end{equation}
This corresponds to contributions to the Wilson coefficients $C_9$ and 
$C_{10}$ at the 
high energy scale given by 
\begin{equation}
\delta C_9=-\delta C_{10} = \frac{\sin^2\theta_W}{2\alpha^2}
\left(\frac{M_W}{m_{\tilde{d}^k_R}}\right)^2\,
\tilde{\lambda}'_{i2k}\tilde{\lambda}'_{i1k}~.
\label{dcs}
\end{equation}
If we now specify $\ell=e$ and use the bounds from Table~\ref{rpvbounds} 
we arrive at the constraint
\begin{equation}
\delta C_9^e=-\delta C_{10}^e \le 1.10\,\left(\frac{\tilde{\lambda}'_{12k}}
{0.04}\right)\,\left(\frac{\tilde{\lambda}'_{11k}}
{0.02}\right)~. 
\label{dce}
\end{equation}
Notice that these are independent of the squark mass, which cancels.
Taking this upper limit on
the Wilson coefficients results in the 
dot-dashed lines of Figs.~\ref{pill} and~\ref{rholl} corresponding to 
$D^+\to\pi^+e^+e^-$ and $D^0\to\rho^0e^+e^-$, respectively. 
The effect in these rates is small, of order $10\%$ at most, 
whereas the experimental bounds are a factor of $20$ above 
this level in the best case (given by the pion mode).

\begin{figure}
\hskip 3.0cm
\epsfig{figure=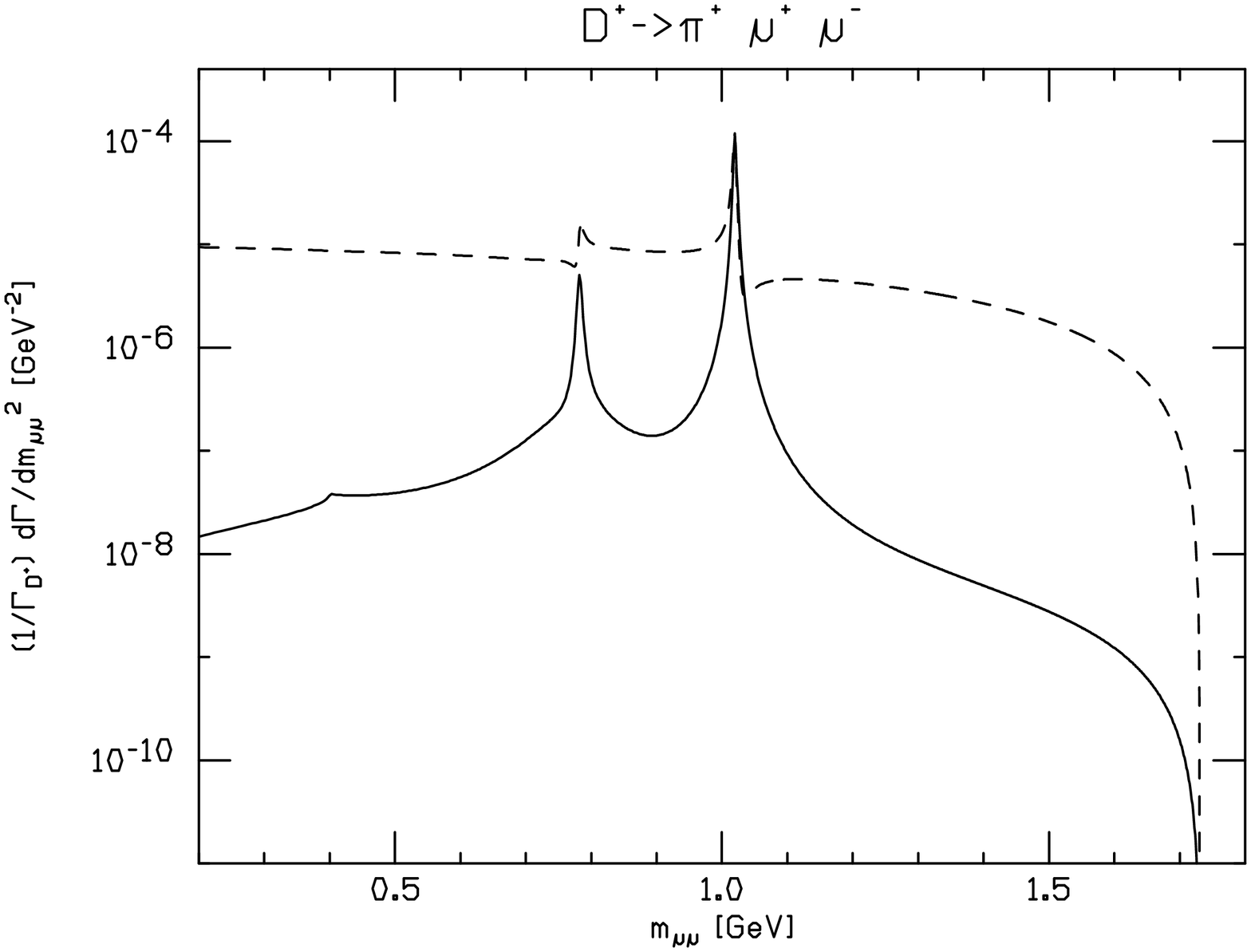,height=4.2in,angle=90}
\caption{The dilepton mass distribution for  
$D^+\to\pi^+ \mu^+\mu^-$ normalized to $\Gamma_{D^+}$. 
The solid line shows the sum of the short and the long distance SM 
contributions.
The dashed 
line includes the allowed R-parity violating contribution from Supersymmetry 
(see text for details).}
\label{pimm}
\end{figure}

\begin{figure}
\hskip 3.0cm
\epsfig{figure=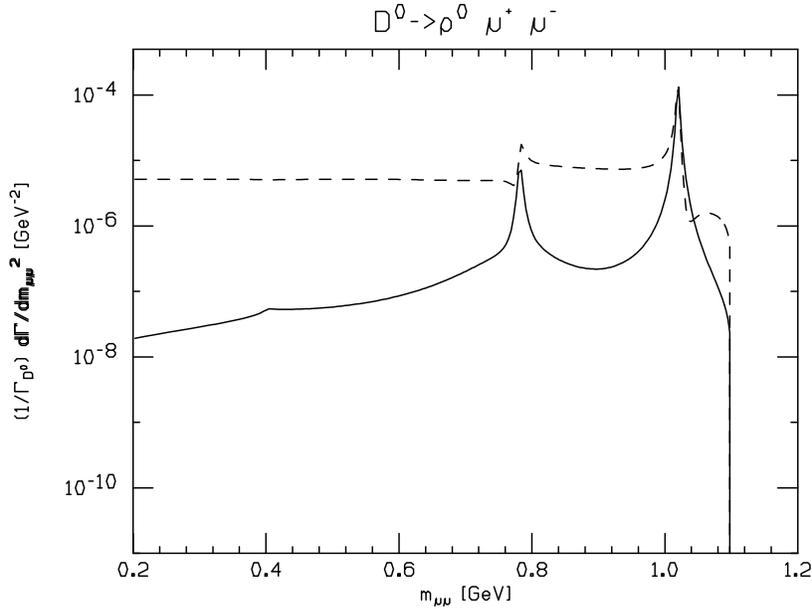,height=4.2in,angle=90}
\caption{The dilepton mass distribution for 
$D^0\to\rho^0 \mu^+\mu^-$ normalized to $\Gamma_{D^0}$. 
The solid line shows the sum of the short and the long distance SM 
contributions.
The dashed line 
includes the allowed R-parity violating contribution from Supersymmetry
(see text for details).}
\label{rhomm}
\end{figure}

On the other hand,   
for $\ell=\mu$ we obtain 
\begin{equation}
\delta C_9^\mu=-\delta C_{10}^\mu \le 17.4\,\left(\frac{\tilde{\lambda}'_{22k}}
{0.21}\right)\,\left(\frac{\tilde{\lambda}'_{21k}}
{0.06}\right)~.
\label{dcmu}
\end{equation}
These upper limits already saturate the experimental bounds 
of ${\cal B}r^{\rm exp}_{D^+\to\pi^+\mu^+\mu^-}<1.5\times10^{-5}$ 
and ${\cal B}r^{\rm exp}_{D^0\to\rho^0\mu^+\mu^-}<2.2\times10^{-5}$ from
Refs.~\cite{e791_1,e791_2}!
Thus we derive the following new constraint on the product of 
R-parity violating couplings,  
\begin{equation}
\tilde{\lambda}'_{22k}\,\tilde{\lambda}'_{21k} < 0.004\ \ , 
\label{newbound}
\end{equation}
which arises from the $D^+\to\pi^+\mu^+\mu^-$ mode.  
This allows for potentially large effects in 
both the $\rho$ and $\pi$ channels as is illustrated in 
Figs.~\ref{pimm} and~\ref{rhomm}.

In Figure~\ref{pimm} we display the dimuon mass distribution 
as a function of the dimuon
mass for $D^+\to\pi^+\mu^+\mu^-$. The solid line, corresponding to the SM
prediction and including both the short and long distance pieces, is clearly
dominated by the latter through the presence of the vector meson resonances
as discussed above.  The dashed line includes the 
contribution of R parity violation, taking 
the R-parity violating coefficients to saturate the above bound in 
Eq.~(\ref{newbound}).  It can be seen that 
{\em away from the resonances} there is an important window for the discovery 
of R parity violation in SUSY theories.
The situation is similar in the $D^0\to\rho^0\mu^+\mu^-$ 
distribution, shown in Figure~\ref{rhomm}. Here, the dashed line is again
obtained by making use of the bound in Eq.~(\ref{newbound}).  This results 
in an upper bound for the 
R parity violating effect given by
${\cal B}r^{R_{\not P}}_{D^0\to\rho^0\mu^+\mu^-}
~<~ 8.7\times10^{-6}$, which is still 
below the experimental limit~\cite{e791_2} 
${\cal B}r^{\rm exp}_{D^0\to\rho^0\mu^+\mu^-} ~<~ 2.2\times10^{-5}$. 

\begin{figure}
\hskip 3.0cm
\epsfig{figure=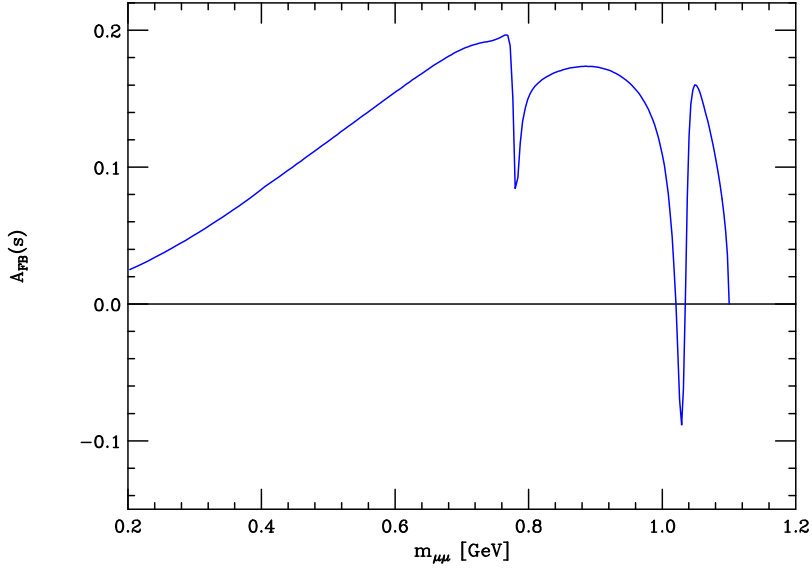,height=4.2in,angle=90}
\caption{The lepton forward-backward asymmetry for 
$D^0\to\rho^0 \mu^+\mu^-$, for the bound of Eq.~(\ref{newbound}). 
(see text for details)}
\label{afbmu}
\end{figure}

In addition to the dilepton mass distribution, this decay mode
also contains angular information as discussed in the previous section. 
For instance, we can define the 
forward-backward asymmetry for leptons as 
\begin{equation}
A_{FB}(q^2)=\frac{
\int_{0}^{1}\frac{d^2\Gamma}{dxdq^2} dx -
\int_{-1}^{0}\frac{d^2\Gamma}{dxdq^2} dx
}
{\frac{d\Gamma}{dq^2}} \ \  ,
\label{afbdef}
\end{equation}
where $x\equiv\cos\theta$, with $\theta$ being the angle between the $\ell^+$ 
and the decaying $D$ meson in the $\ell^+\ell^-$ rest frame.
Expressions for the angular distribution $d\Gamma/dxdq^2$ can be found 
in Ref.~\cite{incas} for the inclusive case and in Ref.~\cite{rareas,otheras} 
for the exclusive modes.
In the SM, $A_{FB}(q^2)$ in $D^0\to\rho^0\ell^+\ell^-$ is negligibly small
throughout the kinematic region.
The reason for this can be seen by inspecting the numerator of the 
asymmetry~\cite{rareas}
\begin{equation}
A_{FB}(q^2) \sim  4\;m_D\;k\;C_{10}
\left\{ C_9^{\rm eff}\;g\;f +\frac{m_c}{q^2}C_7^{\rm eff}
\;\left(f\;G - g\;F\right) \right\}~,
\label{numerator}
\end{equation}
where $k$ is the vector meson three-momentum in the $D$ rest frame, and 
$f$, $g$, $F$ and $G$ are various form-factors.
Since the SM amplitude is dominated by the long distance vector intermediate 
states, we have $C_9^{\rm eff}\gg C_{10}$. New physics contributions that make 
$C_{10}\simeq C_9^{\rm eff}$ will hence 
generate a sizable asymmetry. This is illustrated in the case at hand of
R parity violating supersymmetry. 
For instance, again setting the coupling to the 
values given in Eq.~(\ref{newbound}), we present
the forward-backward asymmetry for 
$D^0\to\rho^0\mu^+\mu^-$ in Figure~\ref{afbmu}.
In order to compute the asymmetry, we make use of 
$D^0\to K^*\ell\nu$ form-factors, together with $SU(3)$ symmetry and 
heavy quark {\em spin} symmetry\footnote{See the first reference 
cited in Ref.~\cite{rareas}.}.  
This gives a bound on the integrated asymmetry of $I_{FB}^{\mu\mu}\simeq 0.15$. 
For $D^0\to\rho^0 e^+ e^-$, we get $I_{FB}^{e e}\simeq 0.08$. 
Supersymmetry could thus produce very sizable asymmetries. In general, 
any non-zero value of $A_{FB}(q^2)$ that is measured should be interpreted
as arising from new physics.

The effective interactions of Eq.~(\ref{dheff1}) 
also lead to a contribution to the two body decay
$D^0\to\mu^+\mu^-$. The R parity violating contribution to the branching
ratio then reads
\begin{equation}
{\cal B}r^{\not R_p}_{D^0\to\mu^+\mu^-} = 
\tau_{D^0}\,f_D^2\,m_\mu^2\,m_D\,\sqrt{1-\frac{4m_\mu^2}{m_D^2}}
\;\frac{\left(\tilde{\lambda}'_{22k}\tilde{\lambda}'_{21k}\right)^2}
{64\pi\,m_{\tilde{d}_k}^4}~.
\label{rpv_d2mu}
\end{equation}
Applying the bound in Eq.~(\ref{newbound}) gives the constraint
\begin{equation}
{\cal B}r^{\not R_p}_{D^0\to\mu^+\mu^-} < 3.5\times 10^{-6}\;
\left(\frac{\tilde{\lambda}'_{12k}}{0.04}\right)^2
\,\left(\frac{\tilde{\lambda}'_{11k}}
{0.02}\right)^2~.
\label{d2mu_bound} 
\end{equation}
The current experimental limit~\cite{e791_1} 
${\cal B}r_{D^0\to\mu^+\mu^-}<5.2\times10^{-6}$ is just above this value, 
implying that future measurements of this
decay mode will constrain the product of these
R parity violating couplings.

Finally, we consider the products of R parity violating couplings 
that lead to lepton flavor violation.
For instance, the products $\tilde{\lambda}'_{11k}\tilde{\lambda}'_{22k}$ and 
$\tilde{\lambda}'_{21k}\tilde{\lambda}'_{12k}$ will give rise to
$D^+\to\pi^+\mu^+ e^-$.
This leads to 
\begin{equation}
\delta C_9^{\mu e}=-\delta C_{10}^{\mu e}=
4.6\times\left\{ \left(\frac{\tilde{\lambda}'_{11k}}{0.02}\right)
\left(\frac{\tilde{\lambda}'_{22k}}{0.21}\right)
 + 
\left(\frac{\tilde{\lambda}'_{21k}}{0.06}\right)
\left(\frac{\tilde{\lambda}'_{12k}}{0.04}\right)
\right\}~,
\label{d2pimue}
\end{equation}
which results in ${\cal B}r^{\not R_p}_{D^+\to\pi^+\mu^+ e^-}
< 3\times10^{-5}$, to be contrasted with~\cite{e791_1} 
${\cal B}r^{\rm exp}_{D^+\to\pi^+\mu^+ e^-} <3.4\times10^{-5}$.
Here again, experiment is on the verge of being sensitive to R parity
violating effects in supersymmetry.  Similarly, for the corresponding 
two body decay we have
\begin{equation}
{\cal B}r^{\not R_p}_{D^0\to\mu^+e^-} <0.5 \times10^{-6}\times
\left\{ \left(\frac{\tilde{\lambda}'_{11k}}{0.02}\right)
\left(\frac{\tilde{\lambda}'_{22k}}{0.21}\right)
 + 
\left(\frac{\tilde{\lambda}'_{21k}}{0.06}\right)
\left(\frac{\tilde{\lambda}'_{12k}}{0.04}\right)
\right\}~,
\label{d2mue_2body}
\end{equation}
whereas the current bound is~\cite{e791_1} 
${\cal B}r^{\rm exp}_{D^0\to\mu^+ e^-} <8.1\times10^{-6}$.
We summarize the results of this section in Table~\ref{rpv_table}.

Finally, we point out that similar effects to those considered in this section
are generated by leptoquarks. Their exchange lead in general to effective 
interactions similar to the $\lambda'$ terms in Eq.~(\ref{rpv}).

\begin{table}[h]
\centering
\begin{tabular}{|l|c|c|c|} \hline\hline
Decay Mode & SM & $\not R_p$ & Expt. Limit\\ \hline
$D^+\to\pi^+e^+e^-$ & $2.0\times10^{-6}$ & $2.3\times10^{-6}$ &
$5.2\times10^{-5}$\\ \hline
$D^0\to\rho^0 e^+ e^-$ & $1.8\times10^{-6}$ & $5.1\times10^{-6}$ 
& $1.0\times10^{-4}$\\ \hline
$D^+\to\pi^+\mu^+\mu^-$ &  $1.9\times10^{-6}$ & $1.5\times10^{-5}$ 
&$ 1.5\times 10^{-5}$\\ \hline
$D^0\to\rho^0 \mu^+ \mu^-$ & $1.8\times10^{-6}$ & $8.7\times10^{-6}$ &
$2.3\times10^{-4}$  \\ \hline
$D^0\to\mu^+\mu^-$ & $3.0\times10^{-13}$ & $3.5\times10^{-6}$ & 
$4.1\times10^{-6}$  \\ \hline
$D^0\to e^+e^-$ & $10^{-23}$ & $1.0\times10^{-10}$ &
$6.2\times10^{-6}$ \\ \hline
$D^0\to\mu^+ e^-$ & $0$ & $1.0\times10^{-6}$ & $8.1\times10^{-6}$  \\ \hline
$D^+\to\pi^+\mu^+e^-$ & $0$ & $3.0\times10^{-5}$  &
$3.4\times10^{-5}$ \\ \hline
$D^0\to\rho^0 \mu^+ e^-$ & $0$ & $1.4\times10^{-5}$ & $4.9\times10^{-5}$\\
\hline
\end{tabular}
\caption
{Comparison of various decay modes between the SM and 
R parity violation. The third column shows how large the R parity
violating effect can be. The experimental limits are from 
Refs.~\cite{PDG00},\cite{e791_1},\cite{e791_2}.
}
\label{rpv_table}
\end{table}

\subsection{Extensions of Standard Model with Extra Higgses, Gauge Bosons,
Fermions, or Dimensions}
 
  In this section we summarize the results from classes of models 
   which have additional Higgs scalar doublets, or family gauge symmetry
   or extra leptons. All of these give rise to flavor changing
   couplings at tree level and hence yield potentially large rates for rare
   decay modes of D mesons.  In addition we briefly discuss the effects
of extra dimensional physics on rare charm transitions.
   
\subsubsection{Multiple Higgs Doublets}

Many extensions of the Standard Model contain more than one Higgs scalar
doublet.  As is well known, this leads in general to tree level FCNC 
couplings and thus decays such as $D^0 \rightarrow \mu^+ \mu^-, e^+ e^-, 
\mu^\pm e^\mp$, {\it etc} may proceed at rates larger than SM
expectations.  In the down quark sector, there are severe constraints on
such couplings from kaon decay modes~\cite{molzon_k}.  This does not necessarily
lead to equally strong constraints on the up-quark sector.  For example,
as was shown long ago~\cite{pakvasa}, it is possible that simple 
symmetries forbid $\Delta S=1$ FCNC without affecting the 
$\Delta C=1$ sector.

Let us write the general effective $\Delta C=1$ interaction as
\begin{equation}
\beta \frac{G_F}{\sqrt{2}} ~ \bar{u} \gamma_5 c ~ \bar{\ell}_1 
  (a + b \gamma_5) \ell_2 \ \ , 
\end{equation}
where $\beta$ is a model dependent dimensionless number, $a$ and $b$
refer to generic scalar and pseudoscalar couplings, respectively, and 
$\ell_1, \ell_2$ refer to the pairs ($\mu, \mu$), ($e, e$) or 
($\mu, e$).
Comparing to the mode $D^+ \rightarrow \mu^+ \nu_\mu$, one can write 
\begin{eqnarray}
{\cal B}r_{D^0 \to \ell_1 \bar{\ell_2}} &\cong& 
\frac{\beta^2}{\mid{U_{cd}\mid^2}} 
\frac{m_D^2}{m_c m_\mu} \frac{a^2 + b^2}{2} 
{\tau_D^+ \over \tau_D^0} ~{\cal B}r_{D^+ \to \mu^+ \nu} 
\nonumber \\
&\cong& 11.35 ~ \beta^2 \frac{a^2 + b^2}{2} \ \ .
\end{eqnarray}
The corresponding branching ratio 
for the three body modes $c \rightarrow u l_i l_j$ is given
by $0.343 \beta^2 \left( a^2 + b^2\right)/2 $.

We have evaluated the parameters $\beta$, $a$ and $b$ 
in several models with multiple Higgs scalar doublets~\cite{pakvasa},\cite{hall}
and computed the branching ratios for rare decay modes of the
$D^0$.  We find that the branching 
ratios for these modes can be as large as
\begin{eqnarray}
{\cal B}r_{D^0\rightarrow\mu^+\mu^-} \sim 8\times 10^{-10}\ , 
\qquad 
{\cal B}r_{D^0\rightarrow e^+e^-} \sim 4\times 10^{-14} \  ,
\qquad 
{\cal B}r_{D^0\rightarrow\mu^\pm e^\mp} \sim 7\times 10^{-10} \ \ , 
\end{eqnarray}
with the corresponding three body modes having branching ratios smaller than 
these by about a factor of $30$.  While still small, these values are
greatly enhanced over those in the SM.


\subsubsection{FCNC in Horizontal Gauge Models}

The gauge sector in the Standard Model has a large global symmetry which
is broken by the Higgs interaction.  By enlarging the Higgs sector, some
subgroup of this symmetry can be imposed on the full SM lagrangian and
break the symmetry spontaneously.  This family symmetry can be global as well
as gauged~\cite{volkov}.  If the new gauge couplings are very weak 
or the gauge boson
masses are large, the difference between a gauged or global symmetry
is rather difficult to distinguish in practice.  In general there would be
FCNC effects from both the gauge and scalar sectors. Here we consider the
gauge contributions.

Let us construct a simple toy model as an example.  Consider a family
symmetry $SU(2)_H$ under which the left-handed quarks (where the superscripts
denote the weak flavor eigenstates) 
\begin{eqnarray*}
\left(
\begin{array}{cc}
u^0 \\
d^0 
\end{array}
\right )_L &
\left (
\begin{array}{c}
c^0 \\
s^0
\end{array} 
\right)_L,
\end{eqnarray*}
and the corresponding left-handed leptons 
\begin{eqnarray*}
\left(
\begin{array}{cc}
\nu^0_{e} \\
e^0 
\end{array}
\right )_L &
\left (
\begin{array}{c}
\nu^0_{\mu} \\
\mu^0
\end{array} 
\right)_L,
\end{eqnarray*}
transform as members of an $I_H = 1/2$ family doublet.  
The third family is assumed to have $I_H =0$.   
The $SU(2)_H$ symmetry
in this model can be  thought of as a
remnant of an $SU(3)_H$ family symmetry which has been broken
to $SU(2)\times U(1)$.
If $\{ G^i_{\mu}\}$ are the gauge fields corresponding to the $SU(2)_H$ and 
we denote $\psi_{d_L^0} = \left (
\begin{array}{c}
d^0 \\
s^0
\end{array}
\right )_L$, 
$\psi_{u_L^0} = \left (
\begin{array}{c}
u^0 \\
c^0
\end{array}
\right )_L$,  {\it etc}, then the gauge interactions are
\begin{equation}
g \left [ \bar{\psi}_{d^{0}_L} \ \gamma_{\mu}
{\bf \tau}\cdot {\bf G}^\mu \psi_{d^{0}_L} \ +  (d^0 \rightarrow u^0) 
\ + (d^0 \rightarrow \ell^0) \right ] \ \ .
\end{equation}
After the symmetry is broken, the mass eigenstate basis is given by
\begin{eqnarray}
\left (
\begin{array}{c}
d \\
s
\end{array}
\right )_L =
U_d
\left (
\begin{array}{c}
d^0 \\
s^0
\end{array}
\right )_L,
\quad
\left (
\begin{array}{c}
u \\
c
\end{array}
\right )_L =
U_u
\left (
\begin{array}{c}
u^0 \\
c^0
\end{array}
\right )_L,
\left (
\begin{array}{c}
e \\
\mu
\end{array}
\right )_L
 =
U_\ell
\left (
\begin{array}{c}
e^0 \\
\mu^0
\end{array}
\right )_L\,.
\end{eqnarray}
The matrices $U_u, U_d$ and $U_\ell$ each contain one angle, $\theta_f$,
and three 
phases. After the symmetry is broken, the three gauge bosons acquire 
different masses, $m_i$.  If the phases are ignored, the matrix
elements for the processes of interest are:
\begin{eqnarray}
{\cal M}_{D^0 \rightarrow \mu^+ \mu^-}  &=& \frac{1}{2} g^2 f_D \ m_\mu
\left [
\frac{\sin \ 2 \theta_u \cos \theta_e}{m_3^2} \ -
\frac{\cos \ 2 \theta_u \sin 2\theta_e}{m_1^2} 
\right ] \bar{\mu} (1+ \gamma_5) {\mu}  \ , \\ \nonumber
{\cal M}_{D^0 \rightarrow e^- \mu^+} &=& \frac{1}{4} g^2 f_D \ m_\mu
\left [
\frac{\cos\ 2 \theta_u \cos 2 \theta_e}{m_1^2} \ +
\frac{1}{m^2_2} \ +
\frac{\sin 2 \theta_u \sin 2 \theta_e}{m_3^2} 
\right ]   \bar{\mu} (1+ \gamma_5) e \ \ .
\end{eqnarray}
Corresponding expressions exist for $K^0$  decay modes, with $\theta_d$ replacing
$\theta_u$.  To proceed further, let us make the simplifying assumption that 
$m_1\approx m_2 \ll m_3$ and that the mixing angles are small.  Then, using 
the constraints from the kaon system, namely the bounds on $K_L \to e \mu$ 
and the known rate for $K_L \rightarrow \mu \bar{\mu}$, we find that 
the branching ratios for charm decay modes can be as large as
\begin{eqnarray}
{\cal B}r_{D^0 \rightarrow \mu^+ \mu^-} \sim 3.10^{-10} \qquad 
{\rm and} \qquad 
{\cal B}r_{D^0 \rightarrow \mu^\pm e^\mp} \sim 2.10^{-13} \ \ ,
\end{eqnarray}
which are enhanced over the SM expectations.

\subsubsection{\bf Extra Fermions} 

Additional fermions beyond those in the three families of the SM 
can contribute to a variety of rare charm decays and can serve to remove
the effective GIM cancellation inherent to these transitions in the SM.
Let us first consider 
the effect of an SU(2) singlet down-type Q=-1/3 quark 
of the kind that occurs in $E_6$ models~\cite{esix}.
This $b'$ quark will contribute in the loop diagrams~\cite{Ref1} which
mediate decays such as $D^0 \rightarrow \mu^+\mu^-$. 
For a mass $m_{b'} \simeq 250$~GeV, the mixing with $u$ and $c$ quarks given
by $\lambda_{b'}=V_{ub'} V^*_{cb'}$ is constrained by the $b'$ contribution to
$\Delta m_D$. With the current bound on $x_D$ $(x_D \equiv \Delta
m_D/\Gamma_D$) of about 3\%~\cite{cleodd}, $\lambda_{b'}$ has to satisfy
$\lambda_{b'} < 0.003$. The $b'$ contribution to $D^0 \rightarrow \mu^+
\mu^-$ can then be of order
\begin{equation}
{\cal B}r_{D^0 \rightarrow \mu^+ \mu^-} (b') \approx 10^{-11} \ \ , 
\end{equation}
which is two orders of magnitude above the SM value.  There will be
similar enhancements for modes such as $D \rightarrow \pi \ell
\bar{\ell}, D \rightarrow \rho\ell \bar{\ell}$ which would be experimentally
detectable.  We note that 
an additional fourth family 
down-type quark belonging to a $SU(2)_L$ doublet would have an identical effect.

\begin{figure}
\vskip .1cm
\hskip 4.0cm
\epsfig{figure=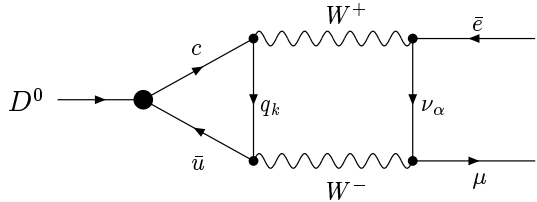,height=1.0in}
\caption{Box diagram mediating $D^0 \to \mu {\bar e}$.} 
\label{fig:dggfigmue}
\end{figure}

When the SM is extended by adding extra lepton doublets or extra
neutral singlets, the decay mode 
$D^0 \to \mu {\bar e}$ can be generated (in a similar fashion as 
$K_L \to \mu {\bar e}$)
only if there are non-degenerate neutrinos 
and nonzero neutrino mixings~\cite{P77}.  We display the relevant 
box-diagram in Fig.~\ref{fig:dggfigmue}.   
The associated matrix element can be written as 
\beq
{\cal M}_{D^0 \to \mu{\bar e}} = {G_F^2 M_W^2 \over 2\pi^2} 
f_D m_\mu B \ {\bar u} \Gamma_{\rm R} v\ \ ,
\label{eu1}
\eeq
where $B$ is given by~\cite{IL81} 
\beqa
B &\equiv& \sum_{\alpha,k} \ U_{\alpha\mu}^* U_{\alpha e} V_{c k}^* 
V_{u k} 
x_\alpha x_k \Bigg[ - {1 \over ( 1 - x_\alpha)(1 - x_k)} 
\nonumber \\
& & \phantom{xxxxx} + {1 \over x_\alpha - x_k} 
\left( {\ln x_k \over (1 - x_k)^2} - 
{\ln x_\alpha \over (1 - x_\alpha)^2} \right) \Bigg] \ \ .
\label{eu2}
\eeqa
In the above, the greek and latin indices run respectively over 
the neutral leptons and negatively-charged quarks, $U_{\alpha\beta}$ 
and $V_{jk}$ are respectively mixing-matrix elements for leptons 
and quarks, 
and $x_k \equiv m_k^2 / M_W^2$.  In the excellent approximation that 
$x_\alpha \simeq 0$ for $\alpha = \nu_e, \nu_\mu, \nu_\tau$ and 
$x_i = 0$ for $i = d$, the expression for $B$ becomes~\cite{LUS88} 
\beqa
B &=&  U_{\mu N} U_{\alpha N}^* \Bigg[ 
V_{cs}^* V_{su} \left( {x_s x_N \over 1 - x_N} - \ln x_s 
+ {\ln x_N \over (1 - x_N)^2} \right) 
\nonumber \\
& & \phantom{xxx} + V_{cb}^* V_{bu} \left( 
{x_b x_N \over 1 - x_N} - \ln x_b 
+ {\ln x_N \over (1 - x_N)^2} \right) \Bigg]  
\nonumber \\
 &\simeq& 4.2 \times 10^{-5} ~U_{Ne}^* U_{N\mu} 
\label{eu3}
\eeqa
for a fourth generation neutral lepton mass of $m_N \simeq 50$~GeV.  
This result varies rather slowly as $m_N$ increases to larger values up to 
and beyond $M_W$.  The decay rate for $D^0 \to \mu {\bar e}$
is then given by 
\beq
\Gamma_{D^0 \to \mu {\bar e}} = \left[ 
{G_F^2 M_W^2 f_D m_\mu B \over 4 \pi^2} \right]^2 {M_D \over 4 \pi} 
\left( U_{Ne}U_{N\mu}\right)^2 \ \ .
\label{eu4}
\eeq
The mixing $\left( U_{Ne}U_{N\mu}\right)^2$ for $m_N > 50$ GeV 
is constrained by the limit on ${\cal B}r_{\mu \to e \gamma}$ 
to be~\cite{AP92,PDG00} less than $5.6 \times 10^{-8}$ and hence we infer
\beq
\Gamma_{D^0 \to \mu {\bar e}} = \left\{ \begin{array}{c}
< 8.62 \times 10^{-27} \ {\rm GeV} \ \ ,\\
\le 1.3 \times 10^{20} \ {\rm sec}^{-1} \ \ .
\end{array} \right. 
\label{eu5}
\eeq
The branching ratio for $D^0 \to \mu {\bar e}$ is thus 
bounded by 
\beq
{\cal B}r_{D^0 \to \mu^- e^+} \le 5.2 \times 10^{-15}  \qquad 
{\rm or} \qquad 
{\cal B}r_{D^0 \to \mu^- e^+ + \mu^+e^-} \le 1.0\times 10^{-14}  
\ \ .
\label{eu6}
\eeq
If the heavy neutral lepton $N^0$ is an $SU(2)$ singlet rather than 
a member of a doublet, the same result is obtained, even though the 
GIM suppression is absent~\cite{LUS88,Gagyi}.  
Hence any 
observation of $D^0 \to \mu {\bar e}$ with 
${\cal B}r_{D^0 \to \mu {\bar e}} > 10^{-14}$ 
cannot be explained by mixing with a heavy neutrino.

\subsubsection{Extra Dimensions}

Attempts to address the hierarchy problem by exploiting the geometry
of space-time have led to extra dimensional theories which have
verifiable consequences at the TeV scale. 
These theories make use of the
idea that our universe lies on a $(3+1)$-dimensional brane which is
embedded in a higher $D$-dimensional space-time, $D\equiv(1+3+\delta)$,
known as the bulk.  The size and geometry of the bulk, as well as the
field content which is allowed to propagate in the bulk,
varies between different scenarios.  Upon compactification of the
additional dimensions, all bulk fields expand into a Kaluza-Klein (KK)
tower of states on the $(3+1)$-brane, where the masses of the KK
states correspond to the $\delta$-dimensional kinetic motion of the
bulk field.  The direct observation or indirect effects of the KK
states signals the existence of extra dimensions.

There are various potential contributions to rare decays within these
scenarios:

(i) In the case of large, flat toroidal extra dimensions~\cite{add},
gravity alone propagates in the bulk and the resultant bulk graviton
KK tower states, $G_n$, couple with inverse Planck scale strength and have very
fine mass splittings given by $1/R_c\sim 10^{-4} eV$ to a few MeV, where
$R_c$ is the common compactification radius of the additional dimensions.
They may be radiated in rare decays such
as $c\to u+G_n$ and subsequently appear as missing energy.  The bulk
graviton KK states couple to the conserved stress-energy tensor, giving
a contribution to this process of order $m_c^2/M_D^2$ where $M_D$ is the
fundamental scale of gravity in the higher dimensional space and is
assumed to be of order a TeV in these models.

(ii) If the extra dimensions are of size TeV$^{-1}$, then the Standard
Model gauge fields may propagate in the bulk and hence expand into KK
towers \cite{ant}.  The KK tower states of the $\gamma, Z, W$, and gluon may
participate in rare transitions in a variety of ways.  However, precision
electroweak data constrain the mass of the first gauge KK excitation to
be in excess of 4 TeV \cite{tgr}, and hence their contributions to rare
decays are small \cite{deshxd}.

(iii)  If the Standard Model fermions are localized \cite{ahs} at specific 
points within a TeV$^{-1}$-sized extra dimension, then they
obtain narrow gaussian-like wave functions in
the extra dimension with a width much smaller than the compactification
radius.  In this case, the fermion mass hierarchy may be explained and
FCNC are suppressed by the small overlap of the
wave functions for the different flavors.

(iv)  The last possibility is the Randall-Sundrum model of localized 
gravity \cite{rs1},  based on a non-factorizable geometry in
5-dimensional Anti-de-Sitter space.  In this case,
the Standard Model gauge and matter fields, as well as gravity, are
allowed to propagate in the warped extra dimension.  The first bulk graviton KK
excitation mass is of order a TeV and hence does not participate in rare
decays.  However, the first gauge and fermion KK excitations are lighter and
may have interesting consequences in rare transitions \cite{dhr3}.  In 
models of this type, it is possible \cite{frank} to generate tree-level FCNC 
which may produce observable effects in rare charm decays.

\subsection{\bf Strong Dynamics }
The possibility that new strong interactions are responsible for 
electroweak symmetry breaking (EWSB) and/or fermion masses has 
important consequences for flavor physics.
The SM with one Higgs doublet already requires the presence of new dynamics at 
a scale $\Lambda$ in order to avoid triviality bounds. The physics
above the cutoff scale gives rise to the scalar sector via bound states
and is connected in some fashion to the the generation of flavor.
For instance, technicolor theories require extended technicolor, whereas 
the generation of the (large) top quark mass may require a 
top-condensation mechanism. In general the generation of fermion 
mass textures leads, in one way or another, 
to FCNC. Here we examine some of the potential effects in rare charm decays and 
their relation to other phenomenological constraints.

\subsubsection{Extended Technicolor}

In standard technicolor theories both fermions and 
techni-fermions transform under the new gauge interaction of
Extended Technicolor (ETC). The condensation of techni-fermions leading to EWSB 
leads to fermion mass terms of the form 
\begin{equation}
m_q\simeq \frac{g^2_{\rm ETC}}{M_{\rm ETC}^2} \langle \bar T
T\rangle_{\rm ETC} \ \ .
\end{equation}
The ETC interactions connect ordinary fermions with techni-fermions, 
as well as fermions and techni-fermions among themselves. The 
relevant sources of FCNC in technicolor models divide into two classes:
those associated with the technicolor sector and those where the diagonal 
ETC gauge bosons acting on ordinary fermions give rise to FCNC through
dimension-six operators. 

The first case gives rise to operators mediated by ETC gauge bosons. 
These in turn have been shown~\cite{css} to give rise to FCNC 
involving the $Z$-boson,  
\begin{eqnarray}
\xi^2\frac{m_c}{8\pi v}\frac{e}{ \sin2\theta_W} U_{cu}^{L} 
Z^\mu~(\bar u_L \gamma_\mu c_L) \quad {\rm and} \quad 
\xi^2\frac{m_t}{8\pi v}\frac{e}{\sin2\theta_W} 
U_{tu}^{L} U_{tc}^{L*}
Z^\mu~(\bar u_L \gamma_\mu c_L) \ \ , 
\label{etc2}
\end{eqnarray}
where $U^L$ is the unitary matrix rotating left-handed up-type 
quark fields into their mass basis and 
$\xi$ is a model-dependent quantity of 
${\cal O}(1)$.
The induced flavor-conserving $Z$ coupling was first studied in 
Ref.~\cite{css} and flavor-changing effects in $B$ decays have been 
examined in Refs.~\cite{rs,gns}.  
The flavor-changing vertices in Eq.~(\ref{etc2}) induce contributions 
to $c \to u\ell^+\ell^-$. These appear mostly as a shift in the 
Wilson coefficient $C_{10}(M_W)$, 
\begin{equation}
  \delta C_{10}\simeq U_{cu}^L\,\frac{m_c}{2v}\, 
\frac{\sin^2\theta_W}{\alpha}\simeq 0.02~,
\label{dc10}
\end{equation}
where we make the assumption $U_{cu}^L\simeq \lambda\simeq 0.22$ 
({\it i.e.}, one power of the Cabibbo
angle) and we take $m_c=1.4~$GeV. Although this represents 
a very large enhancement with respect to the 
SM value of $C_{10}(M_W)$, it does not translate into a large
deviation in the branching ratio.
As mentioned previously, these are dominated by the mixing of 
the operator $O_2$ with $O_9$, leading 
to a very large value of $C_9^{\rm eff}$. The contribution in 
Eq.~(\ref{dc10}) represents only 
a few percent effect in the branching ratio with respect to the SM.
On the other hand, the interaction in Eq.~(\ref{etc2}) can also 
mediate $D^0\to\mu^+\mu^-$. The corresponding
amplitude is 
\begin{equation}
  {\cal A}_{D^0\mu^+\mu^-}\simeq U_{cu}^L\, \frac{m_c}{2\pi v}\, 
\frac{G_F}{\sqrt{2}}\,\sin^2\theta_W
  f_D\,m_\mu~,
  \label{d2mm}
\end{equation}
which should be compared to Eq.~(\ref{d2musm}). This results in the branching 
ratio ${\cal B}r^{\rm ETC}_{D^0\to\mu^+\mu^-}\simeq 0.6\times10^{-10}$,
which although still small, 
is  not only several orders of magnitude larger than the SM short distance
contribution but also more than two orders of magnitude larger than 
the long distance estimates.

Finally, the FCNC vertices of the Z boson in Eq.~(\ref{etc2})
also give large contributions to $c\to u\nu\bar\nu$. The enhancement is 
considerable and results in the branching ratio
\begin{equation}
{\cal B}r^{\rm ETC}_{D^+\to X_u\nu\bar\nu}
\simeq \xi^4\,\left(\frac{U^L_{cu}}{0.2}\right)^2
\,2\times10^{-9}~.
\label{etcnn}
\end{equation}

The second class of contributions from technicolor models comes from the 
diagonal ETC gauge bosons. These generate four-quark interactions 
which refer to a mass scale constrained by $D^0$-$\bar{D^0}$ 
mixing to be approximately $M>100$~TeV~\cite{css}, thus making 
such effects very small in rare charm decays.

\subsubsection{Top-condensation Models}

Top-condensation models postulate
a new gauge interaction that is strong enough to break the top-quark chiral
symmetry and give rise to the large top mass. The various realizations of 
this basic  idea have one common feature: flavor violation. 
Since the new interaction must be non-universal, it mediates FCNC 
at tree level. This arises because the mass matrix generated between 
the top-condensate and the other flavor physics gives rise
to the lighter fermion masses ({\it e.g.} ETC in topcolor-assisted 
technicolor~\cite{tcatc}) and is not aligned with the weak basis. 
Diagonalization of this mass matrix will then leave FCNC vertices 
of the so-called `topcolor interactions' since they couple
preferentially to the third generation. The exchange of top-gluons and
topcolor gauge bosons will generate four-fermion couplings of the form
\begin{eqnarray}
&&\frac{4\pi\alpha_s\cot\theta^2}{M^2} \, U_{tc}^*U_{tu}\,
(\bar u\gamma_\mu T^a t)
(\bar t\gamma^\mu T^a c)
\nonumber\\
&&\frac{4\pi\alpha_s\tan\theta^2}{M^2} \, U_{cu}\,(\bar u\gamma_\mu T^a c)
(\bar c\gamma^\mu T^a c)\nonumber\\
&&\frac{4\pi\alpha_s}{M^2} \, U_{cu}\,(\bar u\gamma_\mu T^a c)
(\bar \xi\gamma^\mu T^a \xi)~,
\label{topc}
\end{eqnarray}
where $\xi^T\equiv (t~b)$, $U_{ij}=U_{ij}^L+U_{ij}^R$ and $M$ is the 
mass of the exchanged color-octet gauge boson.  The first term comes 
from rotating two top-quark fields via the strongly coupled 
topgluon, with the strong interaction being 
reflected in the factor $\cot^2\theta\simeq 22$. The second term
corresponds to a topgluon which is weakly coupled to the first and 
second generations.  In the third term, which  gives the largest 
contribution, the topgluon couples strongly to the third generation 
quark current but weakly to the $(\bar uc)$ current, giving rise to 
a gluon-like coupling. The one-loop insertion of the first and/or 
third terms in Eq.~(\ref{topc}) would result in contributions to the 
operators ${\cal O}_9$ and ${\cal O}_{10}$. However, a term analogous 
to the {\em second} term in Eq.~(\ref{topc}) but with the $\bar c_L$ 
quark rotated to a $\bar u_L$ would contribute to $D^0$-$\bar{D^0}$ 
mixing. The current experimental bound on $\Delta m_D$ taken from 
Eq.~(\ref{ddbound}) implies that 
\begin{equation}
\frac{M}{Re[U_{cu}]} > 140~{\rm TeV}~. 
\label{dd4topc}
\end{equation} 
In standard Topcolor Assisted Technicolor models, this constraint is not
binding on the top-gluon mass since the up-sector rotation matrices 
are taken to be nearly diagonal~\cite{blr}. However, once it is 
satisfied, the bound of Eq.~(\ref{dd4topc}) implies that all effects in rare
charm decays are negligible.  
Similarly, this also applies to the topcolor $Z'$ arising 
from the strongly coupled $U(1)_Y$.

\section{Conclusions}

We have extensively evaluated the potential of 
rare charm decays to probe physics beyond the SM.
In Section~\ref{sec:sm} we computed the SM rates for a variety of decay modes;
incorporating the first evaluation of the QCD corrections to the short
distance contributions, as well as a comprehensive study of long range effects.
This extends our earlier work in Ref.~\cite{bghp}, where we concentrated 
solely on radiative decays. 
We have shown that although, just as in the radiative modes, 
it is still true that long distance contributions dominate the 
rates, there {\it are} decay channels where it is possible to access
the short distance physics. This is particularly true for the case of
$D\to X_u\ell^+\ell^-$ decay modes such as $D\to\pi\ell^+\ell^-$ 
and $D\to\rho\ell^+\ell^-$, away from the resonance
contributions in the low dilepton mass region. This is illustrated
in Figs.~\ref{pill} and~\ref{rholl}, where we see that
for low dilepton invariant mass the sum of long and short distance effects 
leaves a large window where physics beyond the SM can be observed.
Although the uncertainties in our calculation of the long distance 
contributions to this mode are still sizable (roughly of ${\cal O}(1)$)
it is clear that at low dilepton masses new physics effects that are 
an order of magnitude or more larger than the short distance SM signal 
can be detected.  This is not the case in the resonance region where the 
$\phi$, $\omega$ and $\rho$ contributions take the rates to values 
just below current experimental bounds, in a situation analogous 
with radiative decays such as $D\to\rho\gamma$.   
We compile our predictions for the SM rates in Table~\ref{charmbr}.

In Section~\ref{sec:bsm} we explored the potential of these
decays to constrain new physics. In the case of the MSSM, we examined the
sensitivity of rare charm decays to non-universal soft breaking
in the squark mass matrices. We found that large effects are possible
in $D\to\pi\ell^+\ell^-$ and particularly in $D\to\rho\ell^+\ell^-$, as 
can be seen from Figures~\ref{pill_mssm} and~\ref{rholl_mssm}. 
The effect in the vector mode is amplified by the heightened 
sensitivity of this decay channel to the photonic penguin, which carries 
a large enhancement since the gluino helicity flip
replaces the usual charm quark mass insertion.   This effect,
unfortunately, is obscured in radiative decays such as $D\to\rho\gamma$
due to the overwhelming long range effects.  It can therefore, only be
observed by examination of the full dilepton mass spectrum in 
$D\to X\ell^+\ell^-$.
We conclude that an important fraction of parameter space in 
the MSSM with non-universal soft breaking can be 
explored if sensitivities of the order of $10^{-6}$ to $10^{-7}$
in the kinematic region of interest are reached. 

We also considered the effects  of 
R-parity violating couplings in supersymmetry.
We found that the current upper limit on the decay $D\to\pi\mu^+\mu^-$
yields the best constraint on the product 
$\tilde{\lambda}'_{22k}\,\tilde{\lambda}'_{21k}$ (see 
Eq.~(\ref{newbound})). Thus rare charm decays already 
constrain R-parity violating effects!  Our results are summarized in
Table~\ref{rpv_table} for the predictions with 
R-parity violation effects, assuming the couplings saturate their current bounds.
We have also shown that the forward-backward asymmetry for leptons
$A_{FB}$ 
in $D^0\to\rho^0\ell^+\ell^-$ is quite sensitive to these effects 
({\it cf.} Figure~\ref{afbmu}). 
More generally, $A_{FB}$ is negligibly small in the SM due to the 
fact that the vector coupling of leptons is enormously enhanced 
with respect to the axial-vector coupling by the presence of vector 
mesons. Thus, any observation of $A_{FB}$ would point to the presence of
new physics.

We also considered the effects of other non-supersymmetric extensions
of the SM including multi-Higgs models, horizontal gauge models, 
a fourth generation, extra dimensions, as well as models with 
strong dynamics such as extended technicolor and topcolor.
These scenarios give sizeable enhancements in some of the modes.

We conclude that these rare charm decay modes are 
most sensitive to the effects of non-universal supersymmetry 
breaking as well as to R-parity violating couplings. It is then important
to push for increased sensitivity of the experiments, preferably to 
below $10^{-6}$ in order to highly constrain these effects. 
This is in stark contrast with the situation in the radiative 
modes, where sensitivity below $10^{-5}-10^{-6}$ may 
not illuminate short distance physics. 
The dilepton modes should be pursued by all facilities to 
highest possible sensitivity. 

%
\begin{table}
\centering
\begin{tabular}{|l|c|c|c|} \hline\hline
Decay Mode & Experimental Limit & ${\cal B}r_{S.D.}$ & ${\cal B}r_{L.D.}$ 
\\ \hline
$D^+\to X_u^+e^+e^-$ & & $2\times 10^{-8}$ & \\
$D^+\to\pi^+e^+e^-$ & $<4.5\times 10^{-5}$ & &$2\times10^{-6}$ \\
$D^+\to\pi^+\mu^+\mu^-$ & $<1.5\times 10^{-5}$ & &$1.9\times10^{-6}$ \\
$D^+\to\rho^+e^+e^- $ & $<1.0\times 10^{-4}$ & &$4.5\times10^{-6}$ \\
$D^0\to X_u^0e^+e^-$ & & $0.8\times 10^{-8}$ & \\
$D^0\to\pi^0e^+e^-$ & $<6.6\times 10^{-5}$ &  &
$0.8\times10^{-6}$ \\
$D^0\to\rho^0e^+e^-$ & $<5.8\times 10^{-4}$ & & $1.8\times10^{-6} $ \\
$D^0\to\rho^0\mu^+\mu^-$ & $<2.3\times 10^{-4}$ & & $1.8\times10^{-6} $ \\
\hline
$D^+\to X_u^+\nu\bar\nu$ & & $1.2\times 10^{-15}$ & \\
$D^+\to\pi^+\nu\bar\nu$ & &  & $5\times 10^{-16}$ \\
$D^0\to\bar K^0\nu\bar\nu$ &  & & $2.4\times10^{-16}$ \\
$D_s\to \pi^+\nu\bar\nu$ & & & $8\times10^{-15}$ \\ \hline
$D^0\to\gamma\gamma$ & & $3\times10^{-11}$ & few~$\times 10^{-8}$ \\ \hline
$D^0\to\mu^+\mu^-$ & $<3.3\times 10^{-6}$ & $10^{-18}$ &
${\rm few}~\times 10^{-13}$ \\
$D^0\to e^+e^-$ & $<1.3\times 10^{-5}$ & $(2.3-4.7)\times 10^{-24}$ & \\
$D^0\to\mu^\pm e^\mp$ & $<8.1\times 10^{-6}$ & $0$ & $0$ \\ 
$D^+\to\pi^+\mu^\pm e^\mp$ & $<3.4\times 10^{-5}$ & $0$ & $0$ \\ 
$D^0\to\rho^0\mu^\pm e^\mp$ & $<4.9\times 10^{-5}$ & $0$ & $0$ \\ \hline
\end{tabular}
\caption{\small Standard Model predictions for the branching fractions due to 
short and long distance contributions for various rare $D$ meson decays. Also
shown are the current experimental 
limits~\cite{PDG00},\cite{e791_1},\cite{e791_2}.}
\label{charmbr}
\end{table}

\vspace{0.3in} 
{\bf\large Acknowledgments}
The authors acknowledge among others 
Jeff Appel, Dan Kaplan and Boris Kayser for their 
encouragement and motivation, 
Xerxes Tata for helpful discussions, and Jeff Appel and Paul 
Singer for a careful reading of the paper.
The work of G.B was supported by the Director, Office of Science, 
Office of High Energy and Nuclear Physics of the U.S. Department of 
Energy under Contract DE-AC0376SF00098.
The work of E.G. was supported in part by the National 
Science Foundation under Grant PHY-9801875.  The research of
J.H. is supported by the Department of
Energy, Contract DE-AC03-76SF00515.
S.P. was supported in part by the US DOE under grant DE-FG 03-94ER40833.

\eject

\end{document}